\documentclass[aps,superscriptaddress,dvips,11pt]{revtex4}
\usepackage{feynmp}
\usepackage{float}
\usepackage{amsmath}
\usepackage{amsfonts}
\usepackage{amssymb}
\usepackage{epsfig}
\usepackage{graphicx}
\usepackage{subfigure}
\usepackage{bbm}
\usepackage{color}
\usepackage{longtable}
\usepackage{booktabs}
\usepackage{multirow}
\usepackage{graphicx}
\usepackage{epsfig}
\usepackage{bm}
\usepackage{txfonts}
\usepackage{braket}

\newcommand{\Rmnum}[1]{\expandafter\@slowromancap\romannumeral #1@}
\allowdisplaybreaks[3]
\begin{document}

\title{Pion-mediated Cooper pairing of neutrons: beyond the bare vertex
approximation}

\author{Hao-Fu Zhu}
\affiliation{CAS Key Laboratory for Research in Galaxies and
Cosmology, Department of Astronomy, University of Science and
Technology of China, Hefei, Anhui 230026, China}\affiliation{School
of Astronomy and Space Science, University of Science and Technology
of China, Hefei, Anhui 230026, China}
\author{Guo-Zhu Liu}
\altaffiliation{Corresponding author: gzliu@ustc.edu.cn}
\affiliation{Department of Modern Physics, University of Science and
Technology of China, Hefei, Anhui 230026, China}

\begin{abstract}
In some quantum many particle systems, the fermions could form
Cooper pairs by exchanging intermediate bosons. This then drives a
superconducting phase transition or a superfluid transition. Such
transitions should be theoretically investigated by using proper
non-perturbative methods. Here we take the neutron superfluid
transition as an example and study the Cooper pairing of neutrons
mediated by neutral $\pi$-mesons in the low density region of a
neutron matter. We perform a non-perturbative analysis of the
neutron-meson coupling and compute the pairing gap $\Delta$, the
critical density $\rho_{c}$, and the critical temperature $T_c$ by
solving the Dyson-Schwinger equation of the neutron propagator. We
first carry out calculations under the widely used bare vertex
approximation and then incorporate the contribution of the
lowest-order vertex correction. This vertex correction is not
negligible even at low densities and its importance is further
enhanced as the density increases. The transition critical line on
density-temperature plane obtained under the bare vertex
approximation is substantially changed after including the vertex
correction. These results indicate that the vertex corrections play
a significant role and need to be seriously taken into account.
\end{abstract}

\maketitle


\section{Introduction \label{Sec:Introdunction}}

A large part of physical laws, ranging from the fundamental forces
between elementary particles to the emergent phenomena in quantum
many-body systems, can be described by certain types of
fermion-boson interactions \cite{Itzykson}. In the standard model,
quarks and leptons are coupled to a number of intermediate gauge
bosons. In nuclear matter, protons and neutrons interact with each
other by exchanging mesons \cite{Walecka}. In metals, the mutual
influence between itinerant electrons and lattice vibrations is well
captured by an effective electron-phonon coupling \cite{AGD}. If a
fermion-boson coupling is sufficiently weak, one can employ the
perturbation expansion method to compute various physical quantities
\cite{Itzykson, AGD}. However, the perturbation theory becomes
invalid or at least very inaccurate in fermion-boson interacting
models that do not contain any small parameter. It is also
inapplicable when a system undergoes a phase transition. For
instance, in some interacting fermionic systems the fermion-boson
coupling leads to Cooper pairing of fermions, which then drives a
phase transition towards a superconducting phase \cite{BCS} or a
superfluid phase \cite{Migdal60}, depending on whether fermions
carry charges or not. A physically analogous phase transition occurs
in QCD: the strong quark-gluon interaction generates large dynamical
masses for the originally light quarks via the formation of
quark-antiquark pairs \cite{Roberts00}. These phase transitions are
essentially non-perturbative and cannot be investigated by means of
the perturbation expansion method. It is of paramount importance to
develop suitable non-perturbative methods to deal with interacting
fermion-boson systems in which the perturbation theory breaks down.

For fermion-boson systems, the interaction-induced effects on the
single particle properties are embodied in the fully renormalized
fermion and boson propagators, which are classified as two-point
correlation functions in quantum field theory \cite{Itzykson}. The
fermion and boson propagators satisfy two self-consistently coupled
Dyson-Schwinger (DS) integral equations \cite{Itzykson, Roberts00}.
While these two DS equations are exact and absolutely
non-perturbative, they are not self-closed and thus extremely hard
to solve. In these two equations there is an unknown fermion-boson
vertex function. This vertex function is defined in terms of a
three-point correlation function and also satisfies its own DS
equation, which is related to more complicated correlation
functions. In fact, there exist an infinite hierarchy of DS
equations that connect all the $n$-point correlation functions
\cite{Itzykson, Roberts00}. It is certainly not possible to solve
the complete set of DS equations.

The DS equations could be made self-closed if some truncation
schemes are introduced by hands. A widely used truncation is to
simply ignore all the vertex corrections, which amounts to replacing
the full vertex function with the bare vertex. Under such an
approximation, the DS equations of fermion and boson propagators
become self-closed and can be solved to investigate the interaction
effects. The bare vertex approximation has been extensively applied
to study dynamical fermion (e.g., quark) mass generation in gauge
field theories \cite{Roberts00} and to investigate phonon-mediated
superconductivity in various condensed matter systems \cite{Migdal,
Eliashberg, Scalapino}. For the theorists working on relativistic
QED and QCD, the bare vertex approximation is known as the rainbow
approximation \cite{Roberts00}. In condensed-matter community, the
self-closed DS equation of the fermion propagator obtained under the
bare vertex approximation is called Migdal-Eliashberg (ME) equation
\cite{Migdal, Eliashberg, Scalapino}. An obvious fact is that this
truncation scheme is justified only when the vertex corrections are
sufficiently small. In 1958, Migdal \cite{Migdal} made a careful
analysis of the lowest order (one-loop level) vertex correction to
the electron-phonon coupling and argued that its magnitude is
proportional to the factor $\lambda\omega_{D}/E_{F}$, where
$\lambda$ is a dimensional coupling constant, $\omega_{D}$ is
phonons' Debye frequency, and $E_{F}$ is Fermi energy. This factor
is at the order of $0.01 \sim 0.1$ in ordinary metals, thus it is
safe to ignore vertex corrections to the electron-phonon coupling.
Over the last sixty years, the ME equations have been playing a
major role in the theoretical studies of phonon-mediated
superconductivity. It is interesting
to notice that the ME-level DS equations have also been applied to
investigate the properties of phonon-induced nucleon superfluid
pairing in finite nuclei \cite{Terasaki02} and to treat the neutron
pairing due to the coupling of the neutrons to the collective
excitations of crustal lattice based on a Coulomb lattice model of
neutron star crusts \cite{Sedrakian98}. Nevertheless, there exist a
larger number of fermion-boson interacting systems in which the
vertex corrections are not small \cite{Fernandes, Liu21}. To
describe these systems, it is necessary to go beyond the bare vertex
approximation and incorporate vertex corrections into the DS
equations properly.

In this paper, we take the neutron superfluid transition as an
example and study whether the ME equations provide a reliable
description of this transition. The concept of neutron-pairing
induced superfluid was first conceived by Migdal \cite{Migdal60} in
1960, motivated by the microscopic theory of superconductivity
developed by Bardeen, Cooper, and Schrieffer (BCS) in 1957
\cite{BCS}. In nuclear physics it is well established \cite{Walecka}
that nucleons (protons and neutrons) experience an attractive force
at large distances and a repulsive force at small distances. For a
many particle system composed of neutrons, the long-range attraction
binds two neutrons together to form a Cooper pair, similar to the
Cooper pairing of two electrons caused by the phonon-mediated
attraction in superconductors \cite{BCS}. At low temperatures, the
neutron matter enters into a superfluid phase once long-range phase
coherence develops among the Cooper pairs of neutrons.
Interestingly, it has been suggested \cite{Margueron, Mao, Sun10,
Sun, Stein} that superfluid phase would undergo a crossover between
Bose-Einstein condensate (BEC) state and BCS state as the density
becomes sufficiently low. We will not discuss such a crossover in
this paper and focus on the pure superfluid phase. Currently, a
widely accepted notion \cite{Glendenningbook, Shapiro, Pagereview}
is that neutron superfluid exists in the crust and out-core regions
of neutron stars and has an important influence on both the dynamic
and thermal evolutions of neutron stars. Therefore, neutron
superfluid transition deserves a serious theoretical investigation.

Previous theoretical studies on neutron superfluid transition are
dominantly based on the BCS mean-field theory (see review papers
\cite{Lombardoreview, Sedrakian19, Baldo10} and references therein),
using an effective neutron-neutron pairing interaction as the
starting point. Such a pairing interaction is characterized by a
phenomenological potential $V(r)$, which is regarded
\emph{realistic} if it fits the experimental data of scattering
phase-shift with a high precision. A generic agreement
\cite{Lombardoreview, Sedrakian19} is that neutrons undergo a
$^{1}S_{0}$-wave pairing at low densities $\rho$ and a
$^{3}PF_{2}$-wave pairing at higher densities. Although previous
BCS-level studies have made considerable progress, this method has
its own limitations. First, the mean-field treatment neglects many
potentially important effects, such as the retardation of
neutron-neutron interaction and the frequency dependence of various
quantities, and is valid only when the coherence length is large
enough, which is not the case for neutron systems. As pointed out in
Ref.~\cite{Lombardoreview}, BCS mean-field results appear to be
unreliable even in the ultra-low density limit. Second, the BCS
pairing model contains a set of unknown free parameters, which
cannot be fixed when high-precision experimental data are not
available. Indeed, the \emph{realistic} potential used in BCS-level
calculations could be determined by phase shifts only below the
energy scale of roughly $350$MeV \cite{Lombardoreview, Baldo98,
Elgaroy98}. Thus BCS mean-field theory is out of control in neutron
matters (such as neutron star) in which the scattering energies are
too high to be explored by laboratory experiments.

Here we would take a different route to study neutron superfluid
transition. We prefer not to deal with a phenomenological
neutron-neutron potential. Alternatively, we assume that the Cooper
pairing of neutrons results from the Yukawa-type neutron-meson
interactions. In neutron matters there might exist several sorts of
mesons. In order to generate both the long-range attraction and
short-range repulsion, one normally needs to couple neutrons to four
kinds of mesons \cite{Walecka, Glendenningbook, Walecka74, Hirose07,
Kucharek91} including pseudoscalar $\pi$-meson, scalar
$\sigma$-meson, pseudovector $\rho$-meson, and vector
$\omega$-meson. The roles played by these mesons rely on the values
of neutron density, and these mesons cooperate to yield a realistic
neutron-neutron potential. The total Lagrangian would contain four
different neutron-meson interactions, possibly with additional
self-interactions of mesons \cite{Boguta77} and inter-meson
interactions \cite{Muller96}. Although such a model is extremely
complicated and nearly intractable, we believe it important to
investigate this model seriously. On the one hand, the effects
ignored by BCS mean-field theory can be naturally taken into account
within this model. On the other hand, this model provides a
framework to study the superfluid transition and the equation of
states of neutron stars in a unified manner, while BCS mean-field
theory is applicable solely to superfluid transition.

Given that the complicated neutron-meson interacting model is
difficult to handle, we have to break down the daunting task into a
series of simpler steps. We would first find a proper method to
treat a simplified model that contains only one single neutron-meson
interaction and then include other kinds of mesons one by one. In
this paper, we take the first step and consider only one type of
meson. We suppose the neutron density $\rho$ is smaller than
$0.1\rho_0$, where the saturation density is $\rho_{0} \simeq 0.17
\mathrm{fm}^{-3}$, corresponding to relatively large mean neutron
distance $r$. In this region, $\pi$-mesons (i.e., pions) play a
dominant role. Since neutrons do not carry electric charge, they are
only coupled to neutral pion, denoted by $\pi^{0}$. Other mesons,
such as $\sigma$ and $\omega$, must be included at higher densities
($\rho> 0.1\rho_0$) to generate the short-range repulsion, but are
relatively unimportant in the low-density region. We should further
assume that $\rho$ is not too small in order to avoid the possible
BCS-BEC crossover. Although the model describing the Yukawa-type
interaction between neutrons and $\pi^{0}$, hereafter dubbed $\pi N$
interaction, seems to have a simple form, it is a strongly
interacting model because its effective coupling constant $f_{\pi}$
is at the order of unity. Therefore, the perturbative expansion
method is invalid. We will employ the non-perturbative DS equation
approach to investigate the superfluid transition driven by the $\pi
N$ interaction. Different from mean-field calculations, the
retardation effects of $\pi N$ interaction are naturally included in
the DS equations at the outset. Moreover, since the meson propagator
is dynamical, the frequency-momentum dependence of various
quantities can be directly obtained from the DS equation results.

We will first perform a DS equation study of the $\pi N$ model based
on the widely used ME (i.e., bare vertex) approximation. We derive
and solve the self-consistent integral equations of the pairing
function $\Delta_{s}(p)$ and the renormalization function of neutron
energy $A_{0}(p)$ at a series of values of temperature and neutron
density. After solving the coupled equations of $\Delta_{s}(p)$ and
$A_{0}(p)$ numerically, we obtain the energy-momentum dependence of
the $^{1}S_{0}$-wave pairing gap $\Delta(p)=\Delta_{s}(p)/A_{0}(p)$,
extract the critical temperature $T_c$ and the critical density
$\rho_{c}$ of the superfluid transition, and plot a global phase
diagram on the $T$-$\rho$ plane.

Then we examine the impact of vertex corrections on the results
calculated under the bare vertex approximation. At present, it is
hard to compute all the vertex corrections. As the first step, we
consider the one-loop vertex correction and compute its value in the
zero-energy and zero-momentum limits. Our finding is that the
one-loop vertex correction is not negligible even at very low
neutron densities and that its importance is further enhanced as the
density increases. Once the contribution of one-loop vertex
correction is incorporated, the ME results of $\rho_{c}$ and $T_{c}$
are substantially changed. It appears that the ME theory breaks down
and ignoring vertex corrections leads to unreliable results about
the superfluid transition.

The rest of the paper is organized as follows. We define the
effective model of the $\pi N$ interaction in Sec.~\ref{Sec:Model}
and derive the DS integral equations satisfied by the neutron
propagator, the pion propagator, and the $\pi N$ interaction vertex
function in Sec.~\ref{Sec:DSE}. In Sec.~\ref{Sec:MEequations}, we
introduce the bare vertex approximation and decompose the DS
equation of neutron propagator into two self-consistent integral
equations of the neutron pairing function and the renormalization
function of the neutron energy. We then solve the integral equations
to obtain the energy-momentum dependence of these two functions.
Based on the numerical results, we obtain the transition critical
line and show the phase diagram. Next, in
Sec.~\ref{Sec:vertexcorrection} we analyze the relative importance
of the one-loop vertex correction and make a comparison between the
critical lines determined with and without vertex corrections. We
summarize the main results and discuss how to improve the
theoretical analysis of this work in Sec.~\ref{Sec:Summary}. We
provide some calculational details in Appendix \ref{sec:appa}.

\section{Model \label{Sec:Model}}

Although neutron stars were predicted to exist in the universe by
Landau \cite{Shapiro} and independently by Zwicky and Baade
\cite{Baade34} early in 1930s, they had been clearly identified
\cite{Gold68} only three decades later after the group led by Hewish
\cite{Hewish68} observed pulsars. Neutron stars provide a unique
platform to study a variety of intriguing phenomena that are
traditionally investigated separately in astrophysics, particle
physics, nuclear physics, and condensed matter physics. In
particular, it is interesting to explore the neutron superfluid
phase \cite{Migdal60} by analyzing the observational data of neutron
stars. Neutron superfluid is believed to have a significant impact
on both the dynamic and thermal evolutions of neutron stars (for a
recent extensive review, see Ref.~\cite{Pagereview}). The pulsar
glitches, which refers to the abrupt change of spin period, are very
likely due to the relative motion of neutron superfluid to normal
fluid \cite{Baym69}. In addition, some neutron stars, such as
Cassiopeia A \cite{Ho09}, are found to cool down at a unusually high
speed. The underlying mechanism of fast cooling remains unclear,
although a long list of scenarios have been proposed
\cite{Pagereview}. Most of these scenarios rely on the existence of
neutron superfluid \cite{Pagereview}.

To understand the dynamic and thermal evolutions of neutron stars,
it is necessary to determine the conditions for neutron superfluid
to occur and compute some important quantities with high precision,
such as the pairing gap $\Delta(\varepsilon, \mathbf{p})$ at neutron
energies $\varepsilon$ and neutron momenta $\mathbf{p}$, the
critical temperature $T_c$, and the critical neutron density
$\rho_c$. Motivated by this ultimate goal, but apparently without
the ambition of solving the problem immediately, here we study the
pion-mediated Cooper pairing of neutron by using the DS equation
approach. As explained in Sec.~\ref{Sec:Introdunction}, in this
paper we focus on the low neutron density region and consider only
neutral $\pi^{0}$-mesons. The impact of other mesons will be taken
into account in the future.

The Lagrangian density describing the $\pi N$ interaction is of the
form
\begin{eqnarray}
\mathcal{L} &=& \mathcal{L}_n + \mathcal{L}_{\pi^0} +
\mathcal{L}_{{\pi^0} nn} \nonumber \\
&=& \overline{\Psi}_n(x)\left(i\gamma^\mu\partial_\mu -
M_N\right)\Psi_n(x) + \frac{1}{2}\left(\partial_\mu
\phi_0\partial^\mu\phi_0 - m_{\pi}\phi_0^2\right) -i g_{\pi NN}
\overline{\Psi}_n(x)\gamma^5 \Psi_n(x) \phi_0(x).
\label{eq:1}
\end{eqnarray}
Here, $x = (t,\mathbf{x})$ denotes the $(1 + 3)$-dimensional
coordinate vector in real space. $M_N$ is the neutron mass,
$m_{\pi}$ is the $\pi^{0}$ mass, and $g_{\pi NN}$ is the coupling
constant of relativistic $\pi N$ Lagrangian density. $\phi_0$ is a
one-component real scalar field, representing the $\pi^{0}$-meson.
The spinor field $\Psi_n$, whose conjugate is $\overline{\Psi}_n =
\Psi_{n}^{\dag}\gamma^0$, has four components. There are five
$4\times4$ Dirac matrices, defined as follows
\begin{eqnarray}
&&\gamma^0=\begin{bmatrix} \sigma_0 & 0\\0 &
-\sigma_0\end{bmatrix},\gamma^1=\begin{bmatrix} 0 &
\sigma_1\\-\sigma_1 & 0\end{bmatrix}, \gamma^2=\begin{bmatrix} 0 &
\sigma_2\\-\sigma_2 & 0\end{bmatrix}, \gamma^3=\begin{bmatrix} 0 &
\sigma_3\\-\sigma_3 & 0\end{bmatrix}, \nonumber \\
&&\gamma^5=i\gamma^0\gamma^1\gamma^2\gamma^3=\begin{bmatrix} 0 &
\sigma_0\\ \sigma_0 & 0\end{bmatrix},
\end{eqnarray}
where $\sigma_{1}$, $\sigma_{2}$, and $\sigma_{3}$ are standard
Pauli matrices of spin space and $\sigma_0$ is the unit $2\times 2$
matrix. The spinor field $\Psi_n$ satisfies the Dirac equation and
can be expanded as
\begin{eqnarray}
\Psi_n(x)=\sum_{s=\uparrow,\downarrow}\int \frac{d^{3}
\mathbf{p}}{(2\pi)^{3}}\left[\alpha_s(\mathbf{p},t)\mu_s(\mathbf{p})
e^{i\mathbf{p}\cdot\mathbf{x}} + \beta^{\ast}_{s}(\mathbf{p},t)
\nu_s(\mathbf{p})e^{-i\mathbf{p}\cdot\mathbf{x}}\right].
\end{eqnarray}
Here, $\alpha_s(\mathbf{p},t)=\alpha_s(\mathbf{p})e^{-i\varepsilon
t}$, $\beta_s(\mathbf{p},t)=\beta_s(\mathbf{p})e^{-i\varepsilon t}$,
$\mu_s(\mathbf{p}) = N\left[\varphi_s,
\frac{(\boldsymbol{\sigma}\cdot \mathbf{p})}{\varepsilon
+M_N}\varphi_s\right]^T$ and $\nu_s(\mathbf{p}) =
N\left[\frac{(\boldsymbol{\sigma}\cdot\mathbf{p})}{\varepsilon +
M_{N}}\varphi_{-s}, \varphi_{-s}\right]^T$, where $\varepsilon  =
+\sqrt{\mathbf{p}^{2}+M_{N}^{2}}$ and $N=\sqrt{\frac{\varepsilon
+M_N}{2M_{N}}}$, represent the positive and negative frequency
solutions of $\Psi_{n}$, respectively. $\varphi_s$ is the basis of
two-component spin space with subscripts $s = \uparrow,\downarrow$
standing for two spin directions. Defining
$\alpha_s(\mathbf{p},t)=\sqrt{\frac{M_N}{\varepsilon
}}a_s(\mathbf{p})e^{-i\varepsilon t}$ and
$\beta_s(\mathbf{p},t)=\sqrt{\frac{M_N}{\varepsilon
}}b_s(\mathbf{p})e^{-i\varepsilon t}$ for the requirement of second
quantization, we re-write $\Psi_n(x)$ as
\begin{eqnarray}
\Psi_n(x)=\sum_{s=\uparrow,\downarrow}\int
\frac{d^3\mathbf{p}}{(2\pi)^{3}}\sqrt{\frac{M_N}{\varepsilon}}
\left[a_s(\mathbf{p}) \mu_s(\mathbf{p})e^{-ipx} +
b^{\ast}_{s}(\mathbf{p}) \nu_s(\mathbf{p})e^{ipx}\right],
\end{eqnarray}
where $p = (\varepsilon ,\mathbf{p})$ denotes the
$(1+3)$-dimensional momentum vector.

To study the Cooper pairing of neutrons, it is convenient to define
a Nambu spinor \cite{Nambu60} in terms of the neutron field
$\Psi_{n}(x)$. Such a Nambu spinor would have eight components since
$\Psi_{n}$ already has four components. This formal complexity can
be reduced if we consider the non-relativistic limit of the model.
Since the neutron density is supposed to be sufficiently low, the
Fermi momentum is quite small, which allows us to take the
non-relativistic limit. A detailed discussion of the
non-relativistic limit of the originally relativistic $\pi N$ model
is presented in Ref.~\cite{Weise}. In this limit, the anti-neutron
contributions can be omitted and the neutron momentum becomes
relatively unimportant compared to its static mass $(\varepsilon
\approx M_N)$. The field $\Psi_n$ and its conjugate are defined as
\begin{eqnarray}
\Psi_n(x) &=& \sum_{s=\uparrow,\downarrow} \int
\frac{d^3\mathbf{p}}{(2\pi)^{3}}a_{s}(\mathbf{p})
\mu_s(\mathbf{p})e^{-ipx}, \label{eq:5}\\
\overline{\Psi}_n(x) &=& \sum_{s'=\uparrow,\downarrow}\int
\frac{d^3\mathbf{p'}}{(2\pi)^{3}} a^{\ast}_{s'}(\mathbf{p'})
\overline{\mu}_{s'}(\mathbf{p'})e^{ip'x}, \label{eq:6}
\end{eqnarray}
where $\mu_s(\mathbf{p}) = \left[\varphi_{s},\frac{(\boldsymbol{\sigma}
\cdot\mathbf{p})}{2M_N}\varphi_s\right]^T$ and
$\overline{u}_{s'}(\mathbf{p}) = \left[\varphi^\dag_{s'},
-\frac{(\boldsymbol{\sigma}\cdot \mathbf{p'})}{2M_N}
\varphi^{\dag}_{s'}\right]$. Substituting Eq.~(\ref{eq:5}) and
Eq.~(\ref{eq:6}) back to the $\mathcal{L}_{\pi^0nn}$ term of
Lagrangian density (\ref{eq:1}), we obtain
\begin{eqnarray}
\mathcal{L}_{\pi^0nn} &=&-i g_{\pi NN} \sum_{s',s =
\uparrow,\downarrow} \int \frac{d^3\mathbf{p'}
d^3\mathbf{p}}{(2\pi)^6} \left[a^\ast_{s'}(\mathbf{p'})
\overline{\mu}_{s'}(\mathbf{p'}) e^{ip'x}\right]
\gamma^5\left[a_{s}(\mathbf{p})\mu_s(\mathbf{p})
e^{-ipx}\right]\phi_0(q) \nonumber\\
&=& -i g_{\pi NN} \sum_{s',s=\uparrow,\downarrow}\int
\frac{d^3\mathbf{p'}d^3\mathbf{p}}{(2\pi)^6}
a^{\ast}_{s'}(\mathbf{p'}) a_{s}(\mathbf{p})\begin{bmatrix}
\varphi^\dag_{s'} &
-\frac{(\boldsymbol{\sigma}\cdot\mathbf{p'})}{2M_N}\varphi^\dag_{s'}
\end{bmatrix}
\begin{bmatrix}0 & \sigma_0 \\
\sigma_0 & 0\end{bmatrix}\begin{bmatrix} \varphi_s
\\
\frac{(\boldsymbol{\sigma}\cdot\mathbf{p})}{2M_N}\varphi_s\end{bmatrix}
e^{i(p'-p)x}\phi_0(q) \nonumber \\
&=& i g_{\pi NN}\sum_{s',s=\uparrow,\downarrow}\int \frac{d^{3}
\mathbf{p'}d^3\mathbf{p}}{(2\pi)^6}a^\ast_{s'}(\mathbf{p'})
a_{s}(\mathbf{p}) \left[\frac{1}{2M_N}\varphi^{\dag}_{s'}
\boldsymbol{\sigma} \cdot(\mathbf{p'}-\mathbf{p})\varphi_s\right]
e^{i(p'-p)x}\phi_0(q).
\end{eqnarray}
Then we re-define the neutron field operator in the non-relativistic
limit as
\begin{eqnarray}
\psi_n(x) &=& \sum_{s=\uparrow,\downarrow}\int
\frac{d^3\mathbf{p}}{(2\pi)^{3}}
a_{s}(\mathbf{p})\varphi_se^{-ipx}, \label{eq:8}\\
\psi^{\dag}_n(x) &=& \sum_{s'=\uparrow,\downarrow}\int
\frac{d^3\mathbf{p'}}{(2\pi)^{3}}a^{\ast}_{s'}(\mathbf{p'})
\varphi^{\dag}_{s'}e^{ip'x}.\label{eq:9}
\end{eqnarray}
Using Eq.~(\ref{eq:8}) and Eq.~(\ref{eq:9}), the Yukawa-coupling
term is converted into
\begin{eqnarray}
\mathcal{L}_{\pi^0 nn} = -\frac{g_{\pi NN}}{2M_N}
\boldsymbol{\nabla}\cdot \left[\psi_n^{\dag}(x) \boldsymbol{\sigma}
\psi_n(x)\right] \phi_0(x), \label{eq:pinn}
\end{eqnarray}
which is equivalent to
\begin{eqnarray}
\mathcal{L}_{\pi^0 nn} = \frac{f_{\pi}}{m_{\pi}}
\left[\psi_n^{\dag}(x)\boldsymbol{\sigma}\psi_n(x)\right]
\cdot\boldsymbol{\nabla} \phi_0(x).\label{eq:fpi}
\end{eqnarray}
Here, the coupling constants appearing in Eq.~(\ref{eq:pinn}) and
Eq.~(\ref{eq:fpi}) are related to each other by the relation
$\frac{f_{\pi}}{m_{\pi}}=\frac{g_{\pi NN}}{2M_N}$ in
non-relativistic limit \cite{Weise}. The value of the coupling
constant $f_{\pi}$ has already been fixed by experiments
\cite{Weise}: $\frac{f^{2}_{\pi}}{4\pi} \approx 0.08$, or
equivalently $f_{\pi} \approx 1.0$. The neutron mass is $M_N \approx
940 \mathrm{MeV}$ and the pion mass is $m_\pi \approx
135\mathrm{MeV}$. After performing Fourier transformations, we now
obtain the total Lagrangian density in the following
non-relativistic form
\begin{eqnarray}
\mathcal{L} &=& \mathcal{L}_n + \mathcal{L}_{\pi^0} +
\mathcal{L}_{{\pi^0} nn} \nonumber \\
&=& \psi_n^{\dag}(p)\left(\varepsilon-\xi_{\mathbf{p}}\right)
\psi_n(p) + \frac{1}{2}\phi_0^{\dag}(q) \left(\omega^2 -
\mathbf{q}^2-m^2_{\pi}\right)\phi^{}_0(q) +
i\frac{f_{\pi}}{m_{\pi}}\psi_n^{\dag}(p+q)\left(\boldsymbol{\sigma}
\cdot \mathbf{q}\right) \psi_n(p)\phi_0(q).\label{eq:originalmodel}
\end{eqnarray}
In the momentum space, the two-component spinor field $\psi_n(p)$ is
written as $\psi_n(p) = \sum_{s=\uparrow,\downarrow}n_{ps}
\varphi_s=[n_{p\uparrow},n_{p\downarrow}]^T$, where
$\varphi_{\uparrow}=[1, 0]^T$, $\varphi_{\downarrow}=[0, 1]^T$, and
$n_{ps}$ is inversely Fourier transformed from the real-space field
$n_s(x)=\int\frac{d^3\mathbf{p}}{(2\pi)^{3}}a_{s}(\mathbf{p})e^{-ipx}$.
The neutron dispersion now has the form $\xi_{\mathbf{p}} =
\frac{\mathbf{p}^2}{2M_N}-\mu_{n}$, where $\mu_{n}$ is the chemical
potential. The coupling term can be expanded as follows
\begin{eqnarray}
i\frac{f_{\pi}}{m_{\pi}}\psi_n^{\dag}(p+q)
\left(\boldsymbol{\sigma}\cdot\mathbf{q}\right)\psi_n(p)\phi_0(q) &=&
i\frac{f_{\pi}}{m_{\pi}}\Big[q_x\left(n_{p+q\uparrow}^{\dag}
n_{p\downarrow}+n_{p+q\downarrow}^{\dag} n_{p\uparrow}\right)+iq_y
\left(-n_{p+q\uparrow}^{\dag} n_{p\downarrow} +
n_{p+q\downarrow}^{\dag} n_{p\uparrow}\right)
\nonumber \\
&& +q_z\left(n_{p+q\uparrow}^{\dag}
n_{p\uparrow}-n_{p+q\downarrow}^{\dag} n_{p\downarrow}\right)\Big]
\phi_0(q).
\end{eqnarray}
For computational simplicity, we neglect the self-interaction of
meson fields. We notice that a similar model (with an additional
repulsive NN potential) has been studied in Ref.~\cite{Sedrakian03}.

We now define a Nambu spinor \cite{Nambu60} as follows
\begin{eqnarray}
\widetilde{\psi}_n(p) = \left[n_{p\uparrow},n_{p\downarrow},
n^{\dag}_{-p\uparrow},n^{\dag}_{-p\downarrow}\right]^{T}.
\end{eqnarray}
Using this spinor, the total Lagrangian density is re-written as
\begin{eqnarray}
\mathcal{L} &=& \frac{1}{2}\widetilde{\psi}_n^{\dag}(p)
\left[\varepsilon\lambda_0 \otimes\sigma_0 -
\xi_{\mathbf{p}}\lambda_3\otimes\sigma_0\right]
\widetilde{\psi}_n(p) + \frac{1}{2}\phi_0^{\dag}(q)
(\omega^2-\mathbf{q}^2-m^2_{\pi})\phi^{}_0(q) \nonumber \\
&& +i\frac{1}{2}\frac{f_{\pi}}{m_{\pi}}\widetilde{\psi}_n^{\dag}(p+q)
\left[q_x\lambda_0\otimes\sigma_1+q_y\lambda_3\otimes\sigma_2 +
q_z\lambda_0\otimes\sigma_3\right] \widetilde{\psi}_n(p)\phi_0(q).
\end{eqnarray}
Here, $\lambda_{0,1,2,3}$ are $2\times 2$ matrices. The free fermion
(i.e., neutron in Nambu representation) propagator is
\begin{eqnarray}
G_0(p)=\frac{2}{\varepsilon-\xi_{\mathbf{p}}\lambda_3\otimes\sigma_0}
\end{eqnarray}
and the free boson ($\pi^{0}$) propagator is
\begin{eqnarray}
F_0(q) = \frac{1}{\omega^2-\mathbf{q}^2-m^2_{\pi}}.
\end{eqnarray}
The free propagators $G_0(p)$ and $F_0(q)$ are changed by the $\pi
N$ interaction to become renormalized propagators $G(p)$ and $F(q)$,
respectively. The free and renormalized propagators are related to
each other via a set of DS integral equations, which will be
discussed in the next section.

\section{Dyson-Schwinger equations \label{Sec:DSE}}

In quantum field theory, many important physical quantities are
expressed in terms of various $n$-point correlation functions
\cite{Itzykson, Walecka, AGD}, which are defined as the expectation
of the product of $n$ Heisenberg operators $\mathcal{O}_{n}$, namely
\begin{eqnarray}
\langle \mathcal{O}_{1}\mathcal{O}_{2}...\mathcal{O}_{n}\rangle.
\end{eqnarray}
All these correlation functions are connected by a hierarchy of
self-consistent DS equations \cite{Itzykson}. Both the neutron
propagator $G(p)= -i\langle\widetilde{\psi}_n
\widetilde{\psi}_n^{\dag}\rangle$ and the pion propagator
$F(q)=-i\langle \phi^{}_0 \phi_0^{\dag}\rangle$ are two-point
correlation functions. The vertex corrections to $\pi$N coupling are
included in the irreducible vertex function
$\Gamma_{\mathrm{v}}(q,p)$, which is defined via a three-point
correlation function $\langle \phi_0 \widetilde{\psi}_n
\widetilde{\psi}_n^{\dag}\rangle$ as follows
\begin{eqnarray}
F(q)G(p+q)\Gamma_{\mathrm{v}}(q,p)G(p) = \langle \phi_0 \widetilde{\psi}_n
\widetilde{\psi}_n^{\dag}\rangle.\label{eq:EPIvertex}
\end{eqnarray}
As shown in Appendix \ref{sec:appa}, the two propagators satisfy the
following coupled DS equations
\begin{eqnarray}
G^{-1}(p) &=& G_0^{-1}(p)-\frac{1}{2}\frac{f_{\pi}}{m_{\pi}}\int
\frac{d^4q}{(2\pi)^{4}}\widehat{C}_1 G(p+q)F(q)
\Gamma_{\mathrm{v}}(q,p),
\label{eq:DSEGPFULL} \\
F^{-1}(q) &=& F_{0}^{-1}(q)+\frac{1}{2}\frac{f_{\pi}}{m_{\pi}} \int
\frac{d^4p}{(2\pi)^4} \mathrm{Tr}[\widehat{C}_{1} G(p+q)
\Gamma_{\mathrm{v}}(q,p)G(q)], \label{eq:DSEFQFULL}
\end{eqnarray}
where
\begin{eqnarray}
\widehat{C}_{1} = q_x\lambda_0\otimes\sigma_1 + q_y\lambda_3
\otimes\sigma_2+q_z\lambda_0\otimes\sigma_3.
\end{eqnarray}
The fermion-boson interaction vertex function
$\Gamma_{\mathrm{v}}(q,p)$ also satisfies its own DS integral
equation that contains a four-point correlation function. Indeed,
every $n$-point correlation function is related to a $n+1$-point
correlation function via a peculiar DS equation. Apparently, there
exist an infinite number of coupled DS equations since $n$ takes all
the possible positive integers. These DS equations are exact and
contain all the quantum many-body effects produced by the
fermion-boson coupling. In principle, the pairing gap, the $T_c$,
the $\rho_c$ as well as other important quantities could be
simultaneously extracted from the solutions of these DS equations.
Unfortunately, the complete set of DS equations are not closed and
cannot be solved in their original forms.

One route to make the DS equations self-closed is to introduce a
hard truncation. Replacing the full vertex function with the bare
vertex is the simplest and most frequently used truncation scheme
\cite{Roberts00, Eliashberg, Migdal, Scalapino}. Under such an
approximation, the fermion and boson propagators $G(p)$ and $F(q)$
satisfy two self-closed integral equations. Although these two
equations are already greatly simplified, they are still formally
complicated and hard to solve. We notice that there is an
uncertainty about the expression of boson propagator in previous
ME-level studies on superconductivity \cite{Scalapino}: some
theorists simply use the free boson propagator $F_{0}(q)$ to
simplify calculations, while others emphasize the importance of the
boson self-energy and thus prefer to solve the coupled equations of
$F(q)$ and $G(p)$ self-consistently. Below we illustrate that this
uncertainty can be eliminated with the help of an exact relation.

Now let us first define three composite operators
\begin{eqnarray}
j_{\widehat{C}_{i}}(x) = \widetilde{\psi}_n^{\dag}(x)
\widehat{C}_{i}\widetilde{\psi}_n(x),\label{eq:currentoperators}
\end{eqnarray}
where the subscript $i=x,y,z$. Such operators are called current
operators since they have similar forms as various (vector or
scalar) currents. We then use these current operators to define
three current vertex functions $\Gamma_{\widehat{C}_{x,y,z}}$ as
follows
\begin{eqnarray}
\langle j_{\widehat{C}_{i}}(x) \widetilde{\psi}_n(x_1)
\widetilde{\psi}_n^{\dag}(x_2)\rangle = -\int dx_3 dx_4
G(x_1-x_3)\Gamma_{\widehat{C}_{i}}(x,x_3,x_4)G(x_4-x_2),
\label{eq:currentvertex}
\end{eqnarray}
In Appendix \ref{sec:appa}, we show that the interaction vertex
function $\Gamma_{\mathrm{v}}(q,p)$, the current vertex functions
$\Gamma_{\widehat{C}_{x,y,z}}(q,p)$, the full boson propagator
$F(q)$, and the free boson propagator $F_{0}(q)$ are related to each
other via the following exact relation
\begin{eqnarray}
F(q)\Gamma_{\mathrm{v}}(q,p) = -i\frac{1}{2} \frac{f_{\pi}}{m_{\pi}}
F_{0}(q) \left[q_{x} \Gamma_{\widehat{C}_x}(q,p) + q_y
\Gamma_{\widehat{C}_y}(q,p) + q_z\Gamma_{\widehat{C}_z}(q,p)
\right].\label{eq:vertexfunctions}
\end{eqnarray}
Inserting this identity into the DS equation of $G(p)$ leads to
\begin{eqnarray}
G^{-1}(p) = G_0^{-1}(p) + i\frac{1}{4}\left(\frac{f_{\pi}}{m_{\pi}}
\right)^2\int \frac{d^4q}{(2\pi)^4} \widehat{C_1}G(p+q)
F_{0}(q)\left[q_x \Gamma_{\widehat{C}_x}(q,p) + q_y
\Gamma_{\widehat{C}_{y}}(q,p) + q_{z}\Gamma_{\widehat{C}_z}(q,p)
\right].\label{eq:megp}
\end{eqnarray}
Notice that it is $F_{0}(q)$, rather than $F(q)$, that enters into
this equation. If the composite operators defined by
Eq.~(\ref{eq:currentoperators}) are symmetry-induced conserved
currents, the corresponding current vertex functions
$\Gamma_{\widehat{C}_{x,y,z}}(q,p)$ would be connected to the full
fermion propagator $G(p)$ via a number of Ward-Takahashi identities
(WTIs). Every symmetry of the system leads to one specific WTI. As
demonstrated in Refs.~\cite{Liu21, Pan21}, one could prove that the
current vertex functions $\Gamma_{\widehat{C}_{x,y,z}}(q,p)$ depends
only on fermion propagator $G(p)$ if a system contains a sufficient
number of coupled WTIs. For such a system, the DS equation of $G(p)$
would be entirely self-closed and can be solved without introducing
any approximation. In Refs.~\cite{Liu21, Pan21}, it was found that
the electron-phonon interaction and the Coulomb interaction in some
condensed matter systems (metals and semimetals) can be treated in
this non-perturbative manner. Unfortunately, the $\pi N$ interaction
considered in this work is formally much more complicated than
electron-phonon and Coulomb interactions. It is difficult to prove
that $\Gamma_{\widehat{C}_{x,y,z}}(q,p)$ depends solely on the
neutron propagator $G(p)$. At present, we are forced to truncate the
DS equations by hands. In the next two sections, we will first
replace the current vertex functions
$\Gamma_{\widehat{C}_{x,y,z}}(q,p)$ with their bare values, which is
equivalent to the traditional ME approximation, and then examine the
influence of the lowest-order vertex correction to $\pi N$
interaction.

\section{Dyson-Schwinger equation of neutron propagator
under bare vertex approximation \label{Sec:MEequations}}

We first ignore all the quantum (loop-level) corrections to the
current vertex functions $\Gamma_{\widehat{C}_{x,y,z}}(q,p)$ and
make the following replacement
\begin{eqnarray}
q_x \Gamma_{\widehat{C}_x}(q,p) + q_y \Gamma_{\widehat{C}_y}(q,p) +
q_z \Gamma_{\widehat{C}_z}(q,p)\rightarrow -\widehat{C}_{1} =
-q_{x}\lambda_{0}\otimes\sigma_{1}-q_{y}\lambda_{3}\otimes\sigma_{2}
- q_{z}\lambda_{0}\otimes\sigma_{3}.
\end{eqnarray}
On the other hand, the bare interaction vertex function
$\Gamma_{\mathrm{v}}^{\mathrm{bare}}$ takes the form
\begin{eqnarray}
\Gamma_{\mathrm{v}}^{\mathrm{bare}} =
i\frac{1}{2}\frac{f_{\pi}}{m_{\pi}}\widehat{C}_{1}.
\end{eqnarray}
Then the exact relation Eq.~(\ref{eq:vertexfunctions}) requires that
$F(q)\rightarrow F_{0}(q)$. Therefore, we should insert the free
boson propagator $F_{0}(q)$ into the DS equation of $G(p)$ if the
bare vertex approximation is assumed. Some contributions would be
double-counted if the renormalized boson propagator $F(q)$ and the
bare vertex are used simultaneously.

Under the above bare vertex approximation, the DS equation of $G(p)$
becomes
\begin{eqnarray}
G^{-1}(p) = G_0^{-1}(p) - i\frac{1}{4}\left(\frac{f_{\pi}}{m_{\pi}}
\right)^2\int \frac{d^4q}{(2\pi)^4} \widehat{C}_{1}
G(p+q)F_{0}(q)\widehat{C}_{1}.\label{eq:finalme}
\end{eqnarray}
In the region of low neutron density, only $^{1}S_{0}$-wave pairing
is realized \cite{Lombardoreview}. We expand the renormalized
fermion propagator $G(p)$ in the following generic form
\begin{eqnarray}
G(p) = \frac{2}{A_0(p)\varepsilon\lambda_0 \otimes \sigma_0 -
A_1(p)\xi_{\mathbf{p}}\lambda_3\otimes\sigma_0 +
\Delta_s(p)\lambda_1\otimes\sigma_1},\label{eq:genericformgp}
\end{eqnarray}
where $A_0(p)\equiv A_0(\varepsilon, \mathbf{p})$ is the mass
renormalization function, $A_1(p)\equiv A_1(\varepsilon,
\mathbf{p})$ is the chemical potential renormalization, and
$\Delta_s(p)\equiv \Delta_s(\varepsilon, \mathbf{p})$ is the
spin-singlet $^{1}S_{0}$-wave pairing function. The true superfluid
pairing gap is determined by the ratio
$\Delta(p)=\Delta_s(p)/A_0(p)$. As illustrated by Nambu
\cite{Nambu60}, it suffices to suppose a real pairing function
$\Delta_{s}(p)$ since the imaginary part of a complex gap can be
easily obtained from the real part by performing a simple
transformation.

After substituting the generic propagator
Eq.~(\ref{eq:genericformgp}) into the DS equation
(\ref{eq:finalme}), we obtain
\begin{eqnarray}
&& A_0(p)\varepsilon\lambda_0\otimes\sigma_0 -
A_1(p)\xi_{\bm{p}}\lambda_3\otimes\sigma_0 +
\Delta_s(p)\lambda_1\otimes\sigma_1 -
\varepsilon\lambda_0\otimes\sigma_0 +
\xi_{\bm{p}}\lambda_3\otimes\sigma_0 \nonumber \\
&=& -i\left(\frac{f_{\pi}}{m_{\pi}}\right)^{2} \int
\frac{d^4q}{(2\pi)^4}\widehat{C}_{1}
\frac{1}{A_0(p+q)(\varepsilon+\omega) \lambda_0 \otimes \sigma_0 -
A_1(p+q)\xi_{\mathbf{p+q}} \lambda_3\otimes\sigma_0 +
\Delta_s(p+q)\lambda_{1}
\otimes\sigma_1}\widehat{C}_{1}F_{0}(q).\nonumber
\\
\end{eqnarray}
This equation can be readily decomposed into three coupled integral
equations of $A_0(p)$, $A_1(p)$, and $\Delta_s(p)$. Normally,
$A_1(p)$ merely leads to a trivial shift in the chemical potential.
As a good approximation, it is safe to set $A_1(p)=1$. The rest two
renormalization functions $A_0(p)$ and $\Delta_s(p)$ satisfy two
coupled integral equations:
\begin{eqnarray}
A_0(\varepsilon_n,\mathbf{p}) &=& 1+\frac{T}{\varepsilon_n}
\left(\frac{f_{\pi}}{m_{\pi}}\right)^2\sum_{n'} \int \frac{d^3
\mathbf{q}}{(2\pi)^3}\frac{1}{\omega_{n'}^2 +
\mathbf{q}^2+m^2_{\pi}}\nonumber
\\
&& \times \frac{\mathbf{q}^2 A_0(\omega_{n'}+\varepsilon_n,
\mathbf{p+q})(\omega_{n'} + \varepsilon_n)}{A_0^2(\omega_{n'} +
\varepsilon_n,\mathbf{p+q})(\omega_{n'}+\varepsilon_n)^2 +
\xi^2_{\mathbf{p+q}}
+\Delta^2_s(\omega_{n'}+\varepsilon_n,\mathbf{p+q})},\\
\Delta_s(\varepsilon_n,\mathbf{p}) &=&
T\left(\frac{f_{\pi}}{m_{\pi}}\right)^2\sum_{n'} \int \frac{d^3
\mathbf{q}}{(2\pi)^3}\frac{1}{\omega_{n'}^2+\mathbf{q}^2+m^2_{\pi}}
\nonumber \\
&& \times \frac{\mathbf{q}^2\Delta_s(\omega_{n'}+\varepsilon_n,
\mathbf{p+q})}{A_0^2(\omega_{n'}+\varepsilon_n,\mathbf{p+q})(\omega_{n'}
+ \varepsilon_n)^2+
\xi^2_{\mathbf{p+q}}+\Delta^2_s(\omega_{n'}+\varepsilon_n,\mathbf{p+q})}.
\end{eqnarray}
These two equations are self-consistently coupled, manifesting the
mutual influence of Landau damping and Cooper pairing on each other.
By numerically solving them at a series of different values of $T$
and $\rho$, we will be able to obtain the critical temperature $T_c$
and the critical density $\rho_c$ at which the superfluid transition
takes place. The energy-momentum dependence of $A_0(p)$ and
$\Delta_{s}(p)$ can also be simultaneously extracted from the
solutions. The above equations are expressed in the Matsubara
formalism. The fermion frequency is $\varepsilon_{n}=(2n+1)\pi T$
and the boson frequency is $\omega_{n'}=2n'\pi T$, where $n$ and
$n'$ are integers. If we define
$\varepsilon_{n'}=\omega_{n'}+\varepsilon_n$, we find that
$\varepsilon_{n'}$ is restricted to the same region as
$\varepsilon_{n'}=(2{n'}+1)\pi T$. Near the Fermi surface, $A_0(p)$
and $\Delta_s(p)$ are nearly direction independent, which allows us
to suppose that $A_0(\varepsilon_n, \mathbf{p}) = A_0(\varepsilon_n,
|\mathbf{p}|)$ and
$\Delta_s(\varepsilon_n,\mathbf{p})=\Delta_s(\varepsilon_n,
|\mathbf{p}|)$. When the axis of the spherical system is directed
along the vector $\mathbf{p}$, the integration measure becomes
$d^3\mathbf{q} =|\mathbf{q}|^2d|\mathbf{q}| dz d\phi$, where $z =
\cos \theta$ and $\theta$ is the angle between $\mathbf{p}$ and
$\mathbf{q}$. Then the above two equations can be re-written as
follows
\begin{eqnarray}
A_0(\varepsilon_n,|\mathbf{p}|) &=& 1+\frac{T}{\varepsilon_n}
\left(\frac{f_{\pi}}{m_{\pi}}\right)^2\sum_{n'} \int
\frac{|\mathbf{q}|^2d|\mathbf{q}| dz}{(2\pi)^2}
\frac{|\mathbf{p}|^2-2|\mathbf{p}||\mathbf{q}|z +
|\mathbf{q}|^2}{(\varepsilon_{n'} -\varepsilon_n)^2 + |\mathbf{p}|^2
- 2|\mathbf{p}||\mathbf{q}|z +|\mathbf{q}|^2 + m^2_{\pi}} \nonumber
\\
&& \times \frac{A_0(\varepsilon_{n'},|\mathbf{q}|)
\varepsilon_{n'}}{A_0^2(\varepsilon_{n'},|\mathbf{q}|)
\varepsilon_{n'}^2 +
\xi^2_{\bm{q}}+\Delta^2_s(\varepsilon_{n'},|\mathbf{q}|)},
\label{eq:A0} \\
\Delta_s(\varepsilon_n,|\mathbf{p}|) &=&
T\left(\frac{f_{\pi}}{m_{\pi}} \right)^2\sum_{n'} \int
\frac{|\mathbf{q}|^2d|\mathbf{q}| dz}{(2\pi)^2}
\frac{|\mathbf{p}|^2-2|\mathbf{p}||\mathbf{q}|z +
|\mathbf{q}|^2}{(\varepsilon_{n'}-\varepsilon_n)^2 + |\mathbf{p}|^2
-2|\mathbf{p}||\mathbf{q}|z +
|\mathbf{p}|^2 +m^2_{\pi}} \nonumber \\
&& \times \frac{\Delta_s(\varepsilon_{n'},|\mathbf{q}|)}
{A_0^2(\varepsilon_{n'},|\mathbf{q}|)\varepsilon_{n'}^2 +
\xi^2_{\bm{q}} +
\Delta^2_s(\varepsilon_{n'},|\mathbf{q}|)}.\label{eq:Delta}
\end{eqnarray}
These equations are also applicable in the limit of zero
temperature, which is achieved by making the replacement
$T\sum_{n'}\rightarrow \int \frac{d\varepsilon'}{2\pi}$. The
integration range for variable $|\mathbf{q}|$ is set to be
$|\mathbf{q}|\in [0,\Lambda]$, where
$\Lambda$ is an ultraviolet cutoff. The magnitude of $\Lambda$ sets
the highest energy-momentum scale below which the effective $\pi N$
model is valid. Here, we regard $\Lambda$ as a tuning parameter and
choose a suitable $\Lambda$ so as to obtain physically reasonable
results \cite{Kucharek91, Serra01, Hirose07, Sun10} within the
density range (i.e., $\rho \leq 0.1\rho_0$) under consideration. For
neutrons staying on the Fermi surface, the maximal value of
transferred momentum is $|\mathbf{q}|=2p_F$, where $p_F$ is the
Fermi momentum. Thus, we take $\Lambda = 2p_F$. Such a choice of
$\Lambda$ was adopted in a previous investigation of $\pi N$ model
\cite{Sedrakian03}. The density $\rho$ is related to $p_{F}$ via the
relation $\rho = \frac{p_{F}^{3}}{3\pi^2}$, so the cutoff $\Lambda =
2p_F$ is density-dependent. Later we will study the influence of the
variation of $\Lambda$. In principle, the energy $\varepsilon'$
could take all the possible values, namely $\varepsilon' \in
(-\infty,\infty)$. However, we have to introduce a upper bound of
$\varepsilon'$ since we consider only the region of low density. In
practical numerical computations, $\varepsilon'$ is supposed to be
in the range of $(-\Omega,\Omega)$, where the cutoff $\Omega$ should
be sufficiently large. For calculational convenience, we re-scale
all the momenta \cite{Liu21}. For instance, $|\mathbf{q}|$ becomes a
dimensionless variable after it is divided by the cutoff $\Lambda$.
Then the new variable $|\mathbf{q}|$ is defined in the range of
$\in(0, 1)$. Moreover, the re-scale energy now becomes $\varepsilon'
\in(-\frac{\Omega}{\Lambda}, \frac{\Omega}{\Lambda})$. In our
calculations, we choose $\Omega=200$MeV. The model contains only two
turning parameters: the neutron density $\rho$ and the temperature
$T$. The procedure of numerically solving the self-consistent
equations given by Eqs.~(\ref{eq:A0}-\ref{eq:Delta}) can be found in
Ref.~\cite{Liu21}.

\begin{figure}[H]
\centering
\includegraphics[width=3.1in]{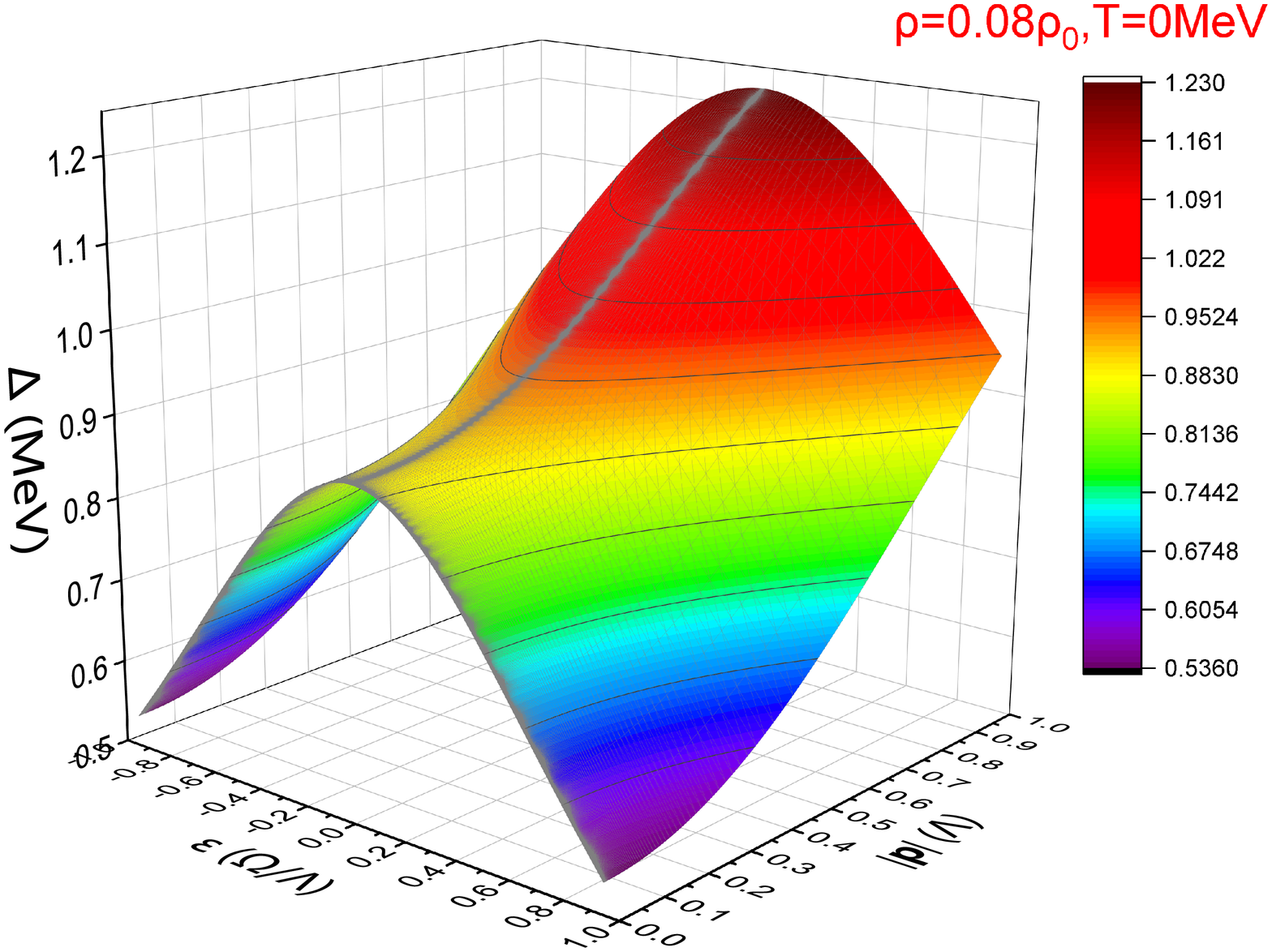}
\includegraphics[width=3.1in]{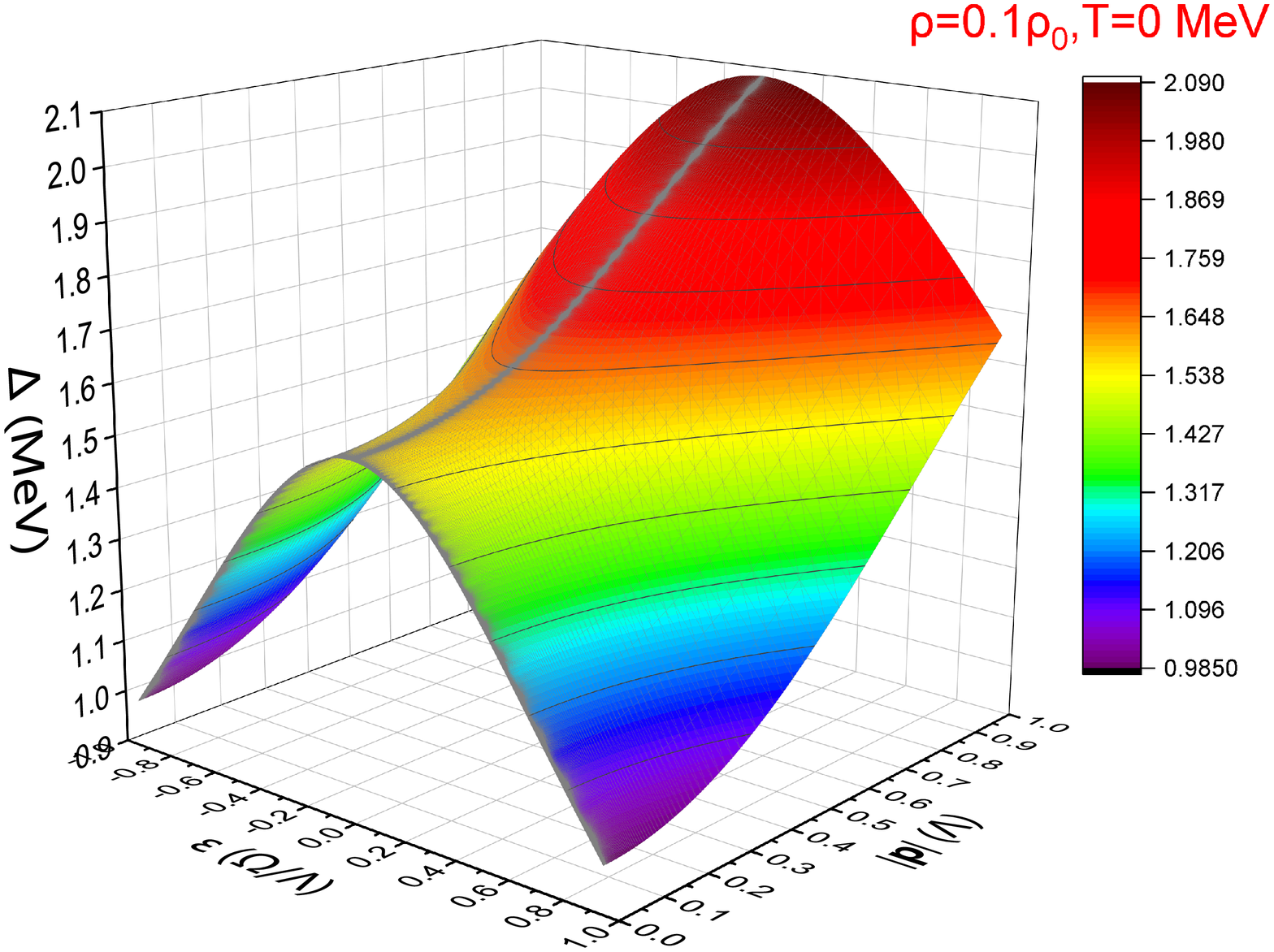}
\includegraphics[width=3.1in]{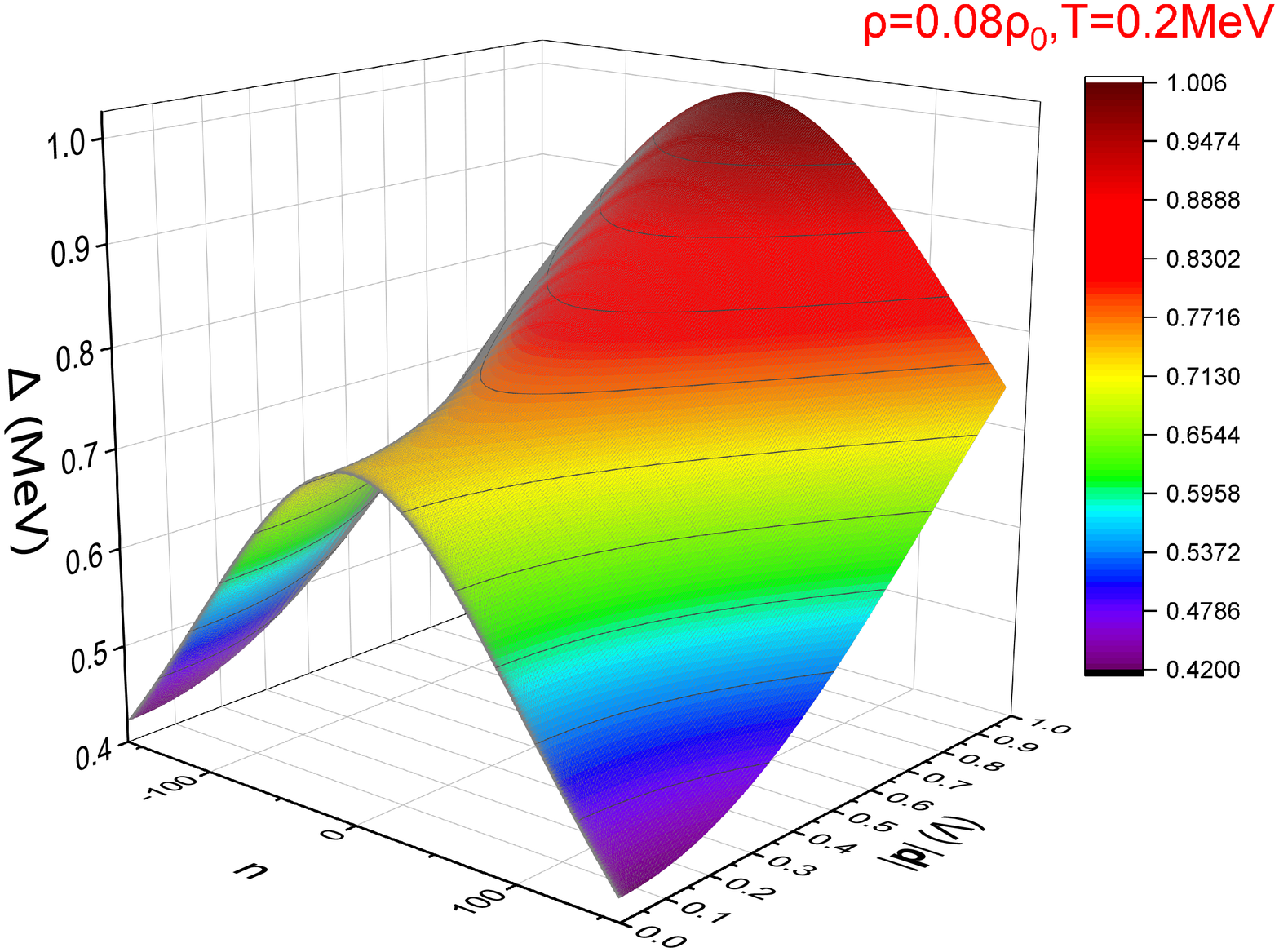}
\includegraphics[width=3.1in]{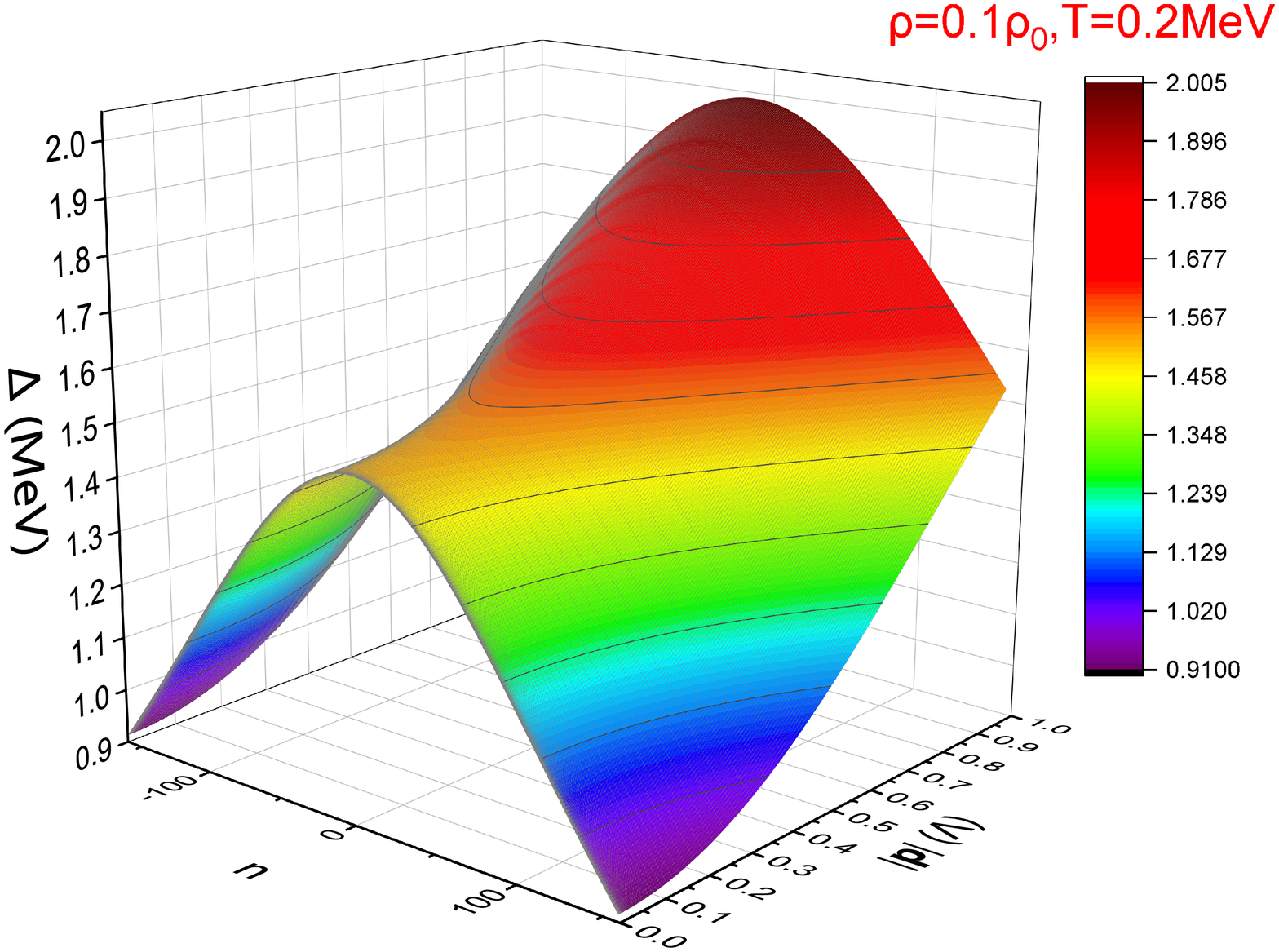}
\includegraphics[width=3.1in]{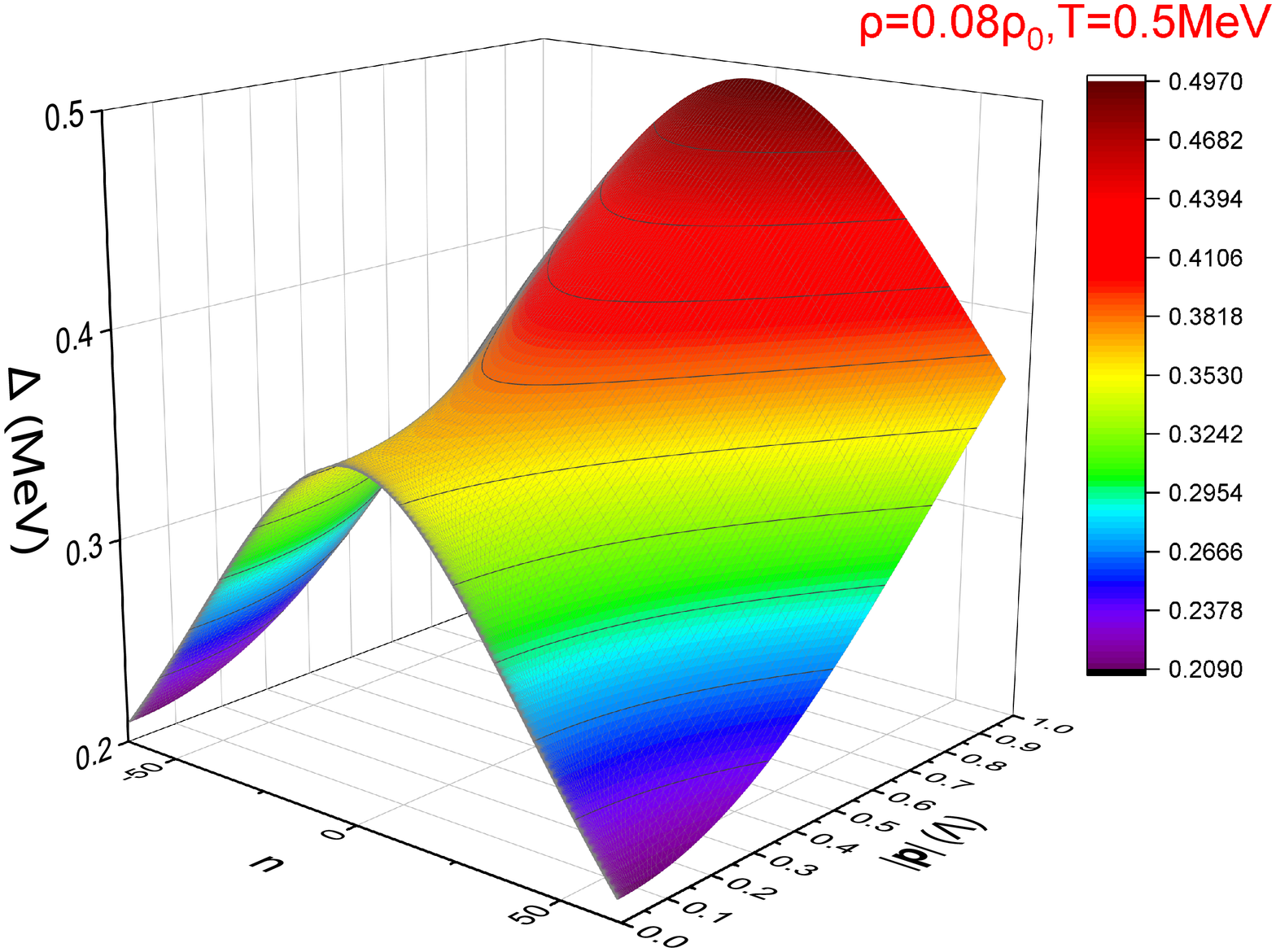}
\includegraphics[width=3.1in]{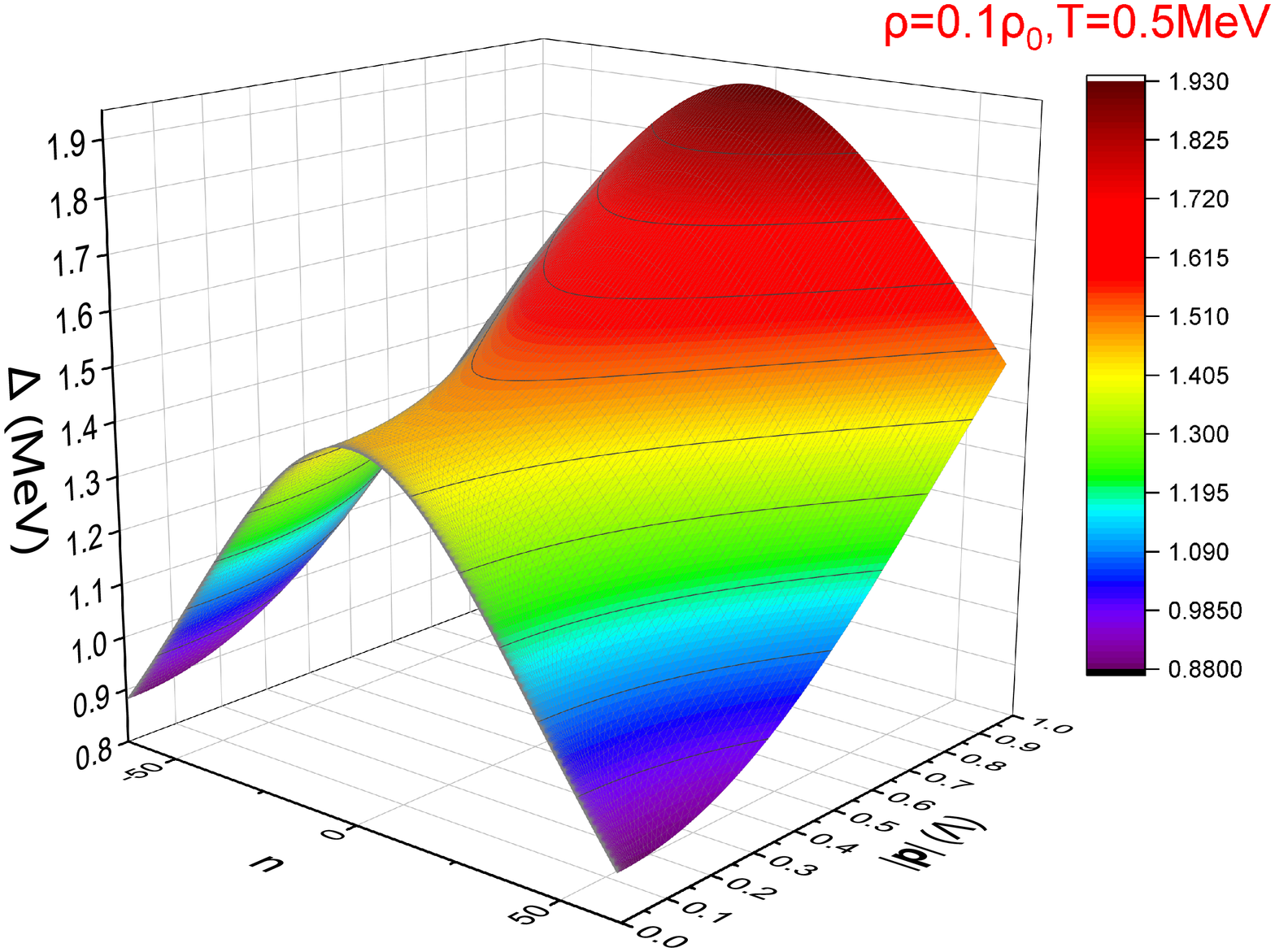}
\caption{The energy-momentum dependence of the pairing gap
$\Delta(\varepsilon,|\mathbf{p}|)$ obtained by solving the
self-consistent integral equations of $\Delta_{s}(\varepsilon,
|\mathbf{p}|)$ and $A_{0}(\varepsilon,|\mathbf{p}|)$. We choose six
different set of tuning parameters $(\rho,T)$. Here and also in
Fig.~\ref{A0figures}, we consider
two representative densities: $\rho=0.08\rho_0$ and
$\rho=0.1\rho_0$. The label $n$ denotes the neutron frequency
$\varepsilon_n=(2n+1)\pi T$ in the case of finite $T$.}
\label{gapfigures}
\end{figure}

In Fig.~\ref{gapfigures}, we present the energy-momentum dependence
of the pairing gap $\Delta(\varepsilon,|\mathbf{p}|)$ at two
representative densities, including $\rho=0.08\rho_{0}$ and
$\rho=0.1\rho_{0}$, and at three different temperatures, including
$T=0$, $T=0.2\mathrm{MeV}$, and $T=0.5 \mathrm{MeV}$. At any given
frequency (energy), $\Delta(|\mathbf{p}|)$ is an increasing function
of $|\mathbf{p}|$. Such a behavior is presumably due to the fact
that the Yukawa coupling is proportional to the transferred momenta
$\mathbf{q}$. At any given $|\mathbf{p}|$, $\Delta(\varepsilon)$
takes its maximal values at zero frequency and decreases as the
absolute value of frequency grows. At a give $T$, the magnitude of
$\Delta$ increases significantly as the density $\rho$ grows. The
maximal value of $\Delta$ is not sensitive on the variation of $T$
at density $\rho = 0.1\rho_0$, but exhibits a much stronger
$T$-dependence at density $\rho = 0.08\rho_0$.

\begin{figure}[H]
\centering
\includegraphics[width=3.1in]{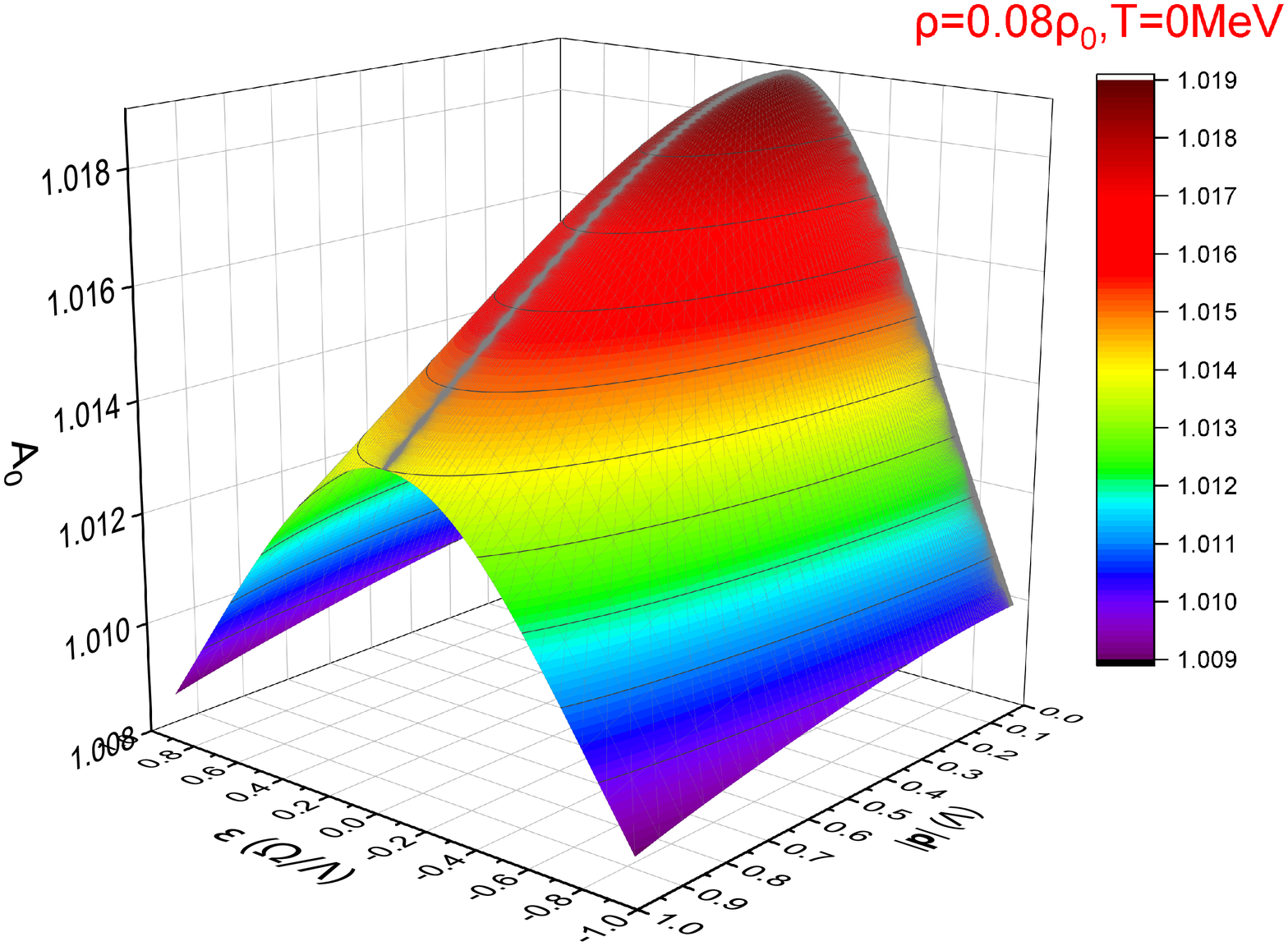}
\includegraphics[width=3.1in]{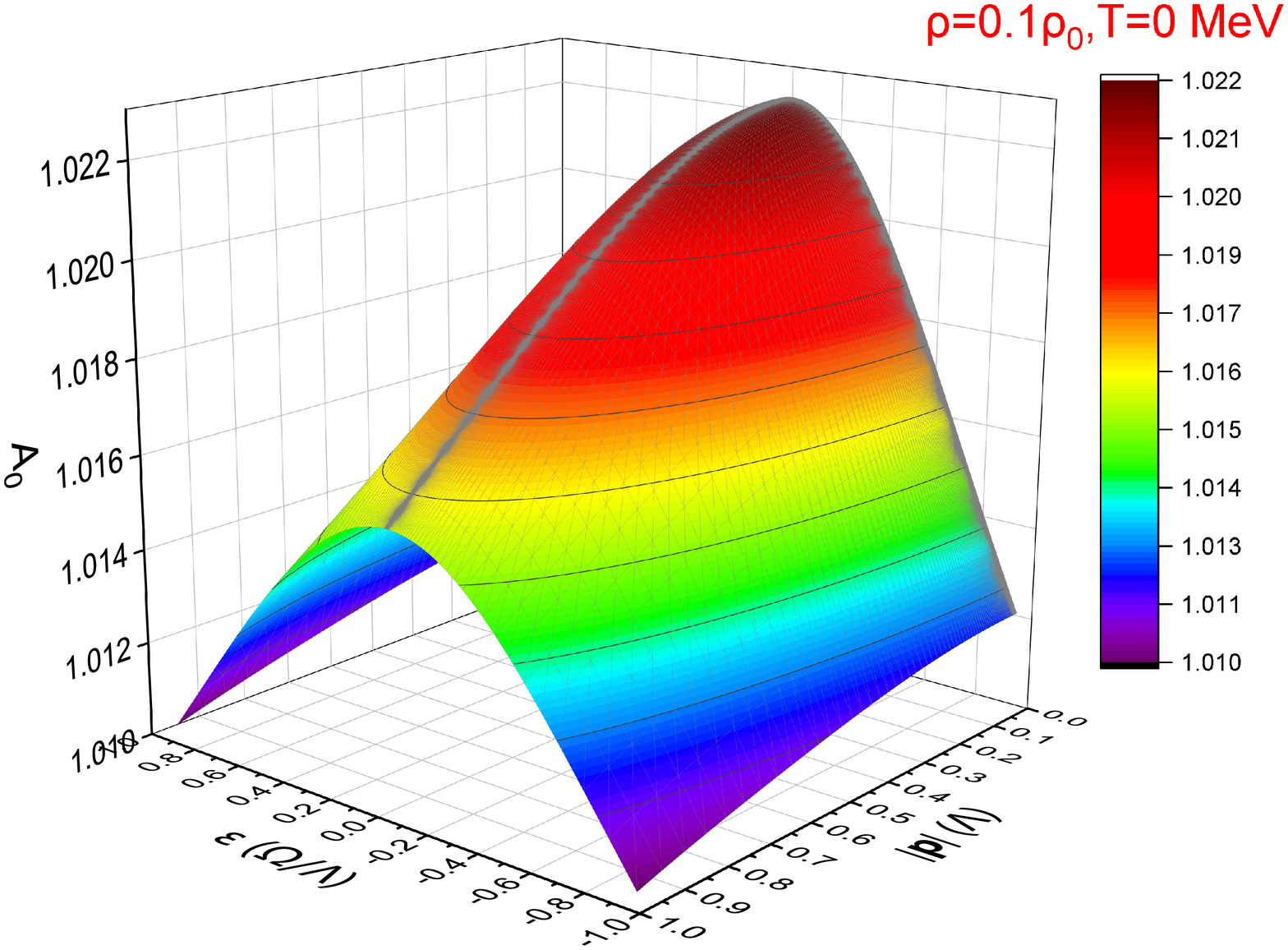}
\caption{The energy-momentum dependence of the renormalization
function $A_0(\varepsilon,|\mathbf{p}|)$ at zero temperature.}
\label{A0figures}
\end{figure}

In Fig.~\ref{A0figures}, we show the energy-momentum dependence of
the renormalization function $A_0(\varepsilon, |\mathbf{p}|)$ at two
densities $\rho=0.08\rho_{0}$ and
$\rho = 0.1\rho_{0}$. The temperature is fixed at $T=0$. Different
from $\Delta(\varepsilon,|\mathbf{p}|)$,
$A_{0}(\varepsilon,|\mathbf{p}|)$ is a decreasing function of
$|\mathbf{p}|$ at any given frequency (energy). At any given
$|\mathbf{p}|$, $A_{0}(\varepsilon,|\mathbf{p}|)$ is largest at zero
frequency, similar to $\Delta(\varepsilon,|\mathbf{p}|)$. The
magnitudes of $A_0$ increases slightly as the density $\rho$ becomes
higher.

\begin{figure}[H]
\centering
\includegraphics[width=3.6in]{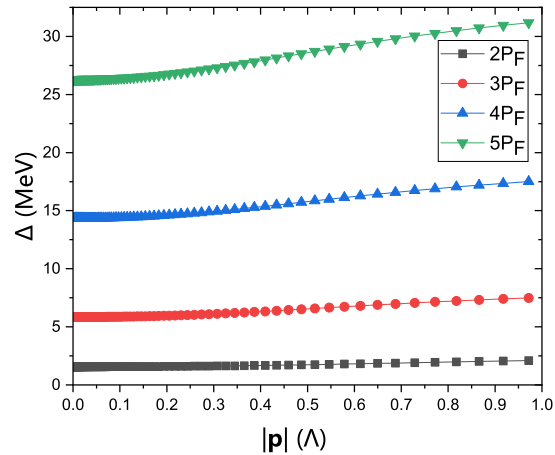}
\caption{The momentum dependence of the pairing gap $\Delta$ at zero
frequency for different values of ultraviolet cutoff $\Lambda$. The
neutron density is $\rho=0.1\rho_0$.} \label{cutoff}
\end{figure}

The above results were obtained by setting $\Lambda = 2p_{F}$. Such
a cutoff appears to be appropriate since it leads to physically
reasonable values of the pairing gap $\Delta$. We have also computed
the pairing gap by adopting other cutoffs and shown the results in
Fig.~\ref{cutoff}. When $\Lambda$ falls within the range $[3 p_F,
5p_F]$, the gap size would be unreasonably large at higher density
$(\rho>0.1\rho_0)$. For instance, as shown in Fig.~\ref{cutoff}, the
gap is as large as $\sim 5 \mathrm{MeV}$ at $\rho = 0.1\rho_0$ if
$\Lambda = 3 p_{F}$. Larger $\Lambda$ leads to significant
enhancement of the pairing gap. From these results, we infer that
$\Lambda = 2p_{F}$ is more suitable than other cutoffs. The strong
cutoff dependence of the gap size should be attributed to the
linear-in-$|\mathbf{q}|$ dependence of the Yukawa coupling between
neutrons and pions. If we consider the coupling of neutrons to other
more massive mesons (such as $\sigma$, $\omega$, and $\rho$) in the
high density region, the pairing gap might exhibit a much weaker
dependence on the values of $\Lambda$. This issue will be addressed
in the future.

\begin{figure}[H]
\centering
\includegraphics[width=3.6in]{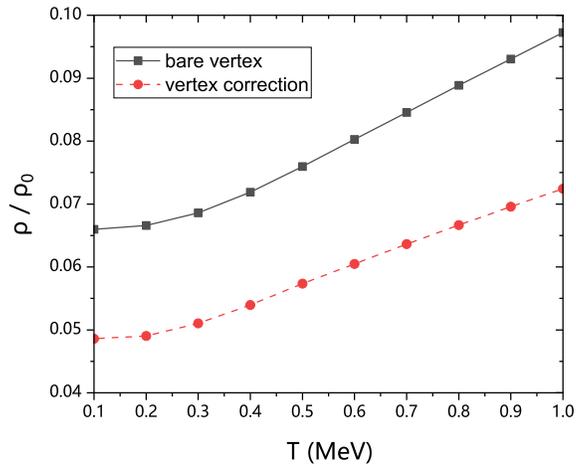}
\caption{Global phase diagram of the
low-density region of neutron matter on the temperature-density
($T$-$\rho$) plane obtained with (dashed line) and without (solid
line) the contribution of one-loop vertex correction. Here, the
density is normalized by saturation density $\rho_0$. The superfluid
phase (upper left corner) and the normal phase (lower right corner)
are separated by a critical line. It is clear that the inclusion of
one-loop vertex correction can substantially alter the critical
line, to be discussed in Sec.~\ref{Sec:vertexcorrection}.}
\label{phasediagram}
\end{figure}

Next we move to calculate the critical values of $\rho_{c}$ and
$T_c$. Such calculations are technically difficult because the
iteration time needed to solve the coupled integral equations
increases dramatically as the critical point is approached. In order
to make numerics simpler, we now set
$A_{0}(\varepsilon,|\mathbf{p}|)=1$ and solve solely the equation of
$\Delta_{s}(\varepsilon,|\mathbf{p}|)$. According to our numerical
results, it is more favorable to realize the neutron superfluid at
higher densities and lower temperatures. In Fig.~\ref{phasediagram},
a global phase diagram is plotted in the plane spanned by the
normalized density $\rho/\rho_{0}$ and the temperature $T$. There is
a critical line between the superfluid phase and the normal, Fermi
liquid phase.

\begin{figure}[H]
\centering
\includegraphics[width=3.6in]{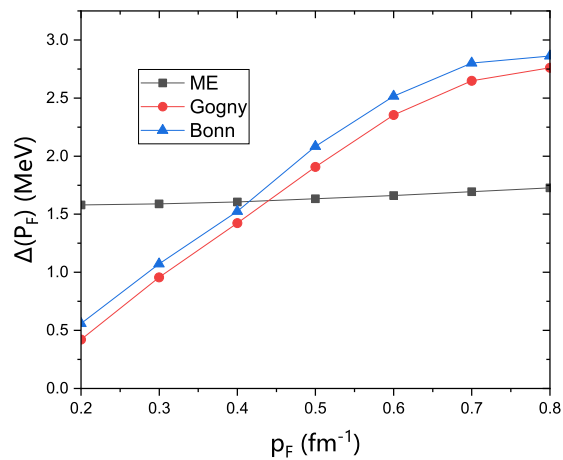}
\caption{The $^{1}S_{0}$-wave neutron pairing gap $\Delta$ as a
function of Fermi momentum $p_F$ obtained under three different
approximations: ME-level, BCS-level based on Gogny force (data are
taken from Refs.~\cite{Kucharek89, Kucharek91, Furtado22} in the
case of $D1$), and BCS-level based on relativistic Bonn potential
(data are taken from Ref.~\cite{Serra01} in the case of version B).}
\label{compare}
\end{figure}

It is now useful to make a comparison between our results and some
previous results on the pairing gap. The $^{1}S_{0}$-wave superfluid
pairing gap $\Delta$ has been previously calculated at the BCS
mean-field level based on a non-relativistic Gogny force $D1$
\cite{Kucharek89, Kucharek91, Furtado22} and also on a relativistic
Bonn potential (so-called version B) \cite{Serra01}. Since both the
long-range attraction and short-range repulsion are incorporated in
the phenomenological Gogny force and Bonn potential, the gap
obtained in these calculations exhibits a non-monotonic $p_{F}$
dependence. It was found \cite{Kucharek89, Kucharek91, Furtado22,
Serra01} that $\Delta(p_{F})$ first increases with growing $p_{F}$
and then decreases with growing $p_{F}$ once $p_{F}$ exceeds roughly
$0.8 \mathrm{fm}^{-1}$, which corresponds to $\rho \approx
0.1\rho_{0}$. In Fig.~\ref{compare}, we compare the $p_F$-dependence
of $\Delta$ obtained by solving the ME equations (\ref{eq:A0}) and
(\ref{eq:Delta}) with the BCS-level results obtained based on the
Gogny force \cite{Kucharek89, Kucharek91, Furtado22} and the Bonn
potential \cite{Serra01} within the range of $0.2 \mathrm{fm}^{-1}
\leq p_{F} \leq 0.8\mathrm{fm}^{-1}$. The Gogny result and Bonn
result are quite similar to each other, but these two results are
substantially different from our ME-level result. The Gogny force
and the Bonn potential are determined by fitting with experimental
data of scattering phase shifts and thus are more realistic than our
$\pi N$ model. However, the BCS mean-field treatment has a serious
limitation: it is carried out by using instantaneous nucleon-nucleon
potentials and entirely ignores the time-dependence of
nucleon-nucleon interaction and the Landau damping effects of
neutrons. From the extensive theoretical studies on phonon-mediated
superconductors \cite{Scalapino}, we already know that the isotopic
effect (the relation between superconducting $T_{c}$ and isotopic
mass) of many metal and alloy superconductors could be well
understood only after the retardation of electron-phonon interaction
and the electron damping are properly considered. In the case of
neutron superfluid, we believe that the retardation of neutron-meson
interaction and the neutron damping are also important and should
not be neglected. Unfortunately, it is difficult to handle these two
effects properly within the framework of BCS mean-field theory. In
comparison, our ME-level calculations take these two effects into
account in a self-consistent manner. The limitation of our present
work is that our results become invalid once $p_{F}$ is larger than
$0.8 \mathrm{fm}^{-1}$ (equivalent to $\rho > 0.1\rho_{0}$) due to
the presence of only one type of meson. In order to investigate the
superfluid transition at higher densities, we will include other
types of mesons, including $\sigma$, $\omega$, and $\rho$, in a
forthcoming ME-level study.

The one-boson-exchange potential is
widely used to study two-nucleon system, nuclear matter, and finite
nuclei \cite{Erkelenz69, Erkelenz74, Holinde87}. The retardation
effect could be partly considered by adding an energy term to the
meson propagator by hands. In principle, one could use such a
potential \cite{Erkelenz71, Holinde72} to derive a BCS-level gap
equation to study superfluid transition. The main merit of this
approach that it contains the contributions from several sorts of
mesons, such as $\pi$, $\rho$, $\delta$, and $\omega$. However, in
this approach only the on-shell processes are included, since the
meson energy is defined by the difference $\omega =
E_{\mathbf{p}'}-E_{\mathbf{p}}$, where $E_{\mathbf{p}} =
\sqrt{\mathbf{p}^2 + M_{N}^{2}}$ is the neutron energy. Moreover, as
demonstrated in Ref.~\cite{Holinde87}, it appears that ignoring the
energy term in the meson propagator is more suitable than keeping
it. Actually, the nucleon-nucleon potential
$V(\mathbf{p},\mathbf{p}')$ and the $R$-matrix
$R(\mathbf{p},\mathbf{p}')$ used to calculate various quantities
\cite{Erkelenz69, Erkelenz74, Holinde87, Erkelenz71, Holinde72}
depend solely on momenta. As a result, the BCS gap equation derived
from such potentials is energy independent, indicating that the
retardation effect is not well treated. Furthermore, the neutron
damping is still not incorporated in such an equation.

\section{Impact of lowest order vertex correction \label{Sec:vertexcorrection}}

The calculations of Sec.~\ref{Sec:MEequations} are based on the
approximation of neglecting all the vertex corrections. Such an
approximation, usually phrased as Migdal theorem \cite{Migdal,
Scalapino, Fernandes} in the condensed-matter community, has been
broadly adopted in previous ME studies. The Migdal theorem is
believed to be well justified in the case of weak electron-phonon
interactions in ordinary metals with a large Fermi surface
\cite{Migdal, Scalapino, Fernandes}. In this section, we examine
whether such a theorem is applicable in the $\pi N$ model.

\begin{figure}[H]
\centering
\includegraphics[width=4.4in]{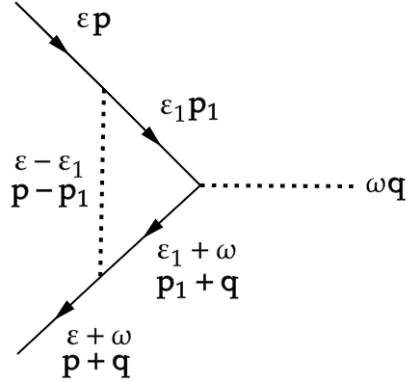}
\caption{The Feynman diagram of one-loop vertex correction. The
solid line represents the free fermion (electron or neutron)
propagator and the dotted line represents the free boson (phonon or
$\pi^{0}$) propagator.} \label{oneloopdiagram}
\end{figure}

To determine the importance of the electron-phonon vertex
corrections, Migdal \cite{Migdal} calculated the lowest order vertex
correction, which is represented by the one-loop diagram shown in
Fig.~\ref{oneloopdiagram}. For ordinary acoustic phonons, there is a
characteristic Debye frequency $\omega_{D}$ that serves as an upper
cutoff of phonon frequencies. This allows one to retain solely the
contributions from the fixed frequency $\omega_{D}$. Making use of
such an approximation, Migdal \cite{Migdal} found that the one-loop
vertex correction is proportional to $\sim \lambda
\omega_{D}/E_{F}$. For ordinary metals, $\lambda \ll 1$ and $E_{F}
\gg \omega_{D}$. In some element metals \cite{Fernandes}, including
Hg, Pb, and Al, the ratio $\omega_{D}/E_{F}$ is as small as $\sim
10^{-2}$. It is absolutely safe to omit all the electron-phonon
vertex corrections for these systems. However, the $\pi N$ model is
very different from electron-phonon systems. First, as
aforementioned the coupling constant $f_{\pi} \approx 1.0$.
Apparently, the $\pi N$ interaction is much stronger than
electron-phonon interactions. Second, the $\pi N$ interacting system
does not have such a characteristic energy scale as Debye frequency.
It appears that the $\pi N$ vertex corrections cannot be suppressed
by any small parameter.

Now we make a more quantitative analysis of the one-loop vertex
correction to $\pi N$ interaction, following the strategy of Migdal
\cite{Migdal}. To avoid unnecessary formal complications, we do not
utilize the Nambu spinor but turn to consider the Lagrangian density
given by Eq.~(\ref{eq:originalmodel}). The one-loop $\pi N$ vertex
correction can be expressed as
\begin{eqnarray}
\Gamma_{1} &\propto& \left(\frac{f_{\pi}}{m_{\pi}}\right)
\left(\boldsymbol{\sigma}\cdot\mathbf{q}\right)
\left(\frac{f_{\pi}}{m_{\pi}}\right)^2\int \frac{d\varepsilon_1
d^3\mathbf{p_1}}{(2\pi)^4}
\left[\boldsymbol{\sigma}\cdot\left(\mathbf{p-p_1}\right)\right]
\left[\boldsymbol{\sigma}\cdot\left(\mathbf{p-p_1}\right)\right]
\nonumber \\
&& \times D_0(\varepsilon-\varepsilon_1,\mathbf{p-p_1})
G_0(\varepsilon_1,\mathbf{p_1}) G_0(\varepsilon_1 +
\omega,\mathbf{p_1+q})
\nonumber \\
&=& -\left(\frac{f_{\pi}}{m_{\pi}}\right)\left(\boldsymbol{\sigma}
\cdot\mathbf{q}\right) \left(\frac{f_{\pi}}{m_{\pi}}\right)^2
\int\frac{d\varepsilon_1 d^3\mathbf{p_1}}{(2\pi)^4} \nonumber \\
&& \times \frac{\left(\mathbf{p-p_{1}}\right)^2}{\left((\varepsilon
- \varepsilon_1)^{2} + (\mathbf{p-p_{1}})^2 +
m^2_{\pi}\right)\left(i\varepsilon_1 -
\xi_{\mathbf{p_1}}\right)\left(i(\varepsilon_1+\omega) -
\xi_{\mathbf{p_1+q}}\right)}.
\end{eqnarray}
Here, the free fermion and boson propagators are given by
$G_0(\varepsilon,\mathbf{p}) = \frac{1}{i\varepsilon -
\xi_{\mathbf{p}}}$ and $D_0(\omega,\mathbf{q}) =
-\frac{1}{\omega^2+\mathbf{q}^2+m^2_{\pi}}$, respectively.

\begin{figure}[H]
\centering
\includegraphics[width=3.6in]{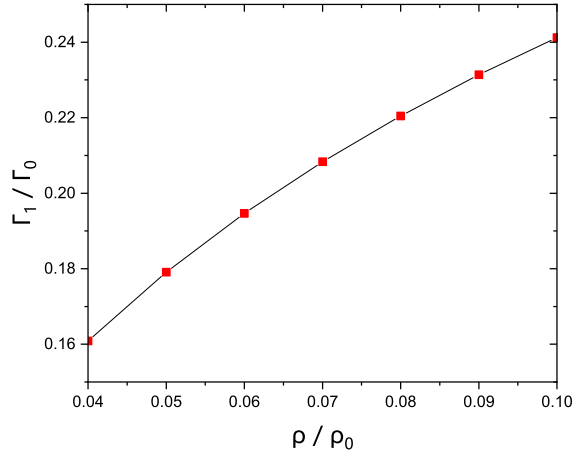}
\caption{The ratio between one-loop vertex correction and bare
vertex as a function of the relative neutron density
$\rho/\rho_{0}$. The one-loop vertex correction becomes more
important as the neutron density increases.} \label{ratio}
\end{figure}

The importance of one-loop vertex correction can be characterized by
the ratio $\Gamma_{1}/\Gamma_{0}$, where
\begin{eqnarray}
\Gamma_{0}\sim \left(\frac{f_{\pi}}{m_{\pi}}\right)
\left(\boldsymbol{\sigma}\cdot\mathbf{q}\right)
\end{eqnarray}
is the bare vertex. This ratio has the following expression
\begin{eqnarray}
\frac{\Gamma_{1}}{\Gamma_{0}} \propto - \left(
\frac{f_{\pi}}{m_{\pi}}\right)^2 \int \frac{d\varepsilon_{1}
d^{3}\mathbf{p_1}}{(2\pi)^4} \frac{\left(\mathbf{p-p_{1}}
\right)^2}{\left((\varepsilon - \varepsilon_1)^2 +
(\mathbf{p-p_{1}})^2+m^2_{\pi}\right) \left(i\varepsilon_1 -
\xi_{\mathbf{p_1}}\right) \left(i\left(\varepsilon_1 + \omega\right)
- \xi_{\mathbf{p_1+q}}\right)},
\end{eqnarray}
which can be further written as
\begin{eqnarray}
\frac{\Gamma_{1}}{\Gamma_{0}} \propto
\left(\frac{f_{\pi}}{m_{\pi}}\right)^2\int \frac{d\varepsilon_1
|\mathbf{p_1}|^2 d|\mathbf{p_1}| dz }{(2\pi)^3}
\frac{\left(|\mathbf{p}|^2-2|\mathbf{p}||\mathbf{p_1}|z +
|\mathbf{p_1}|^2\right) \left(\varepsilon_1
\left(\varepsilon_1+\omega\right) - i\varepsilon_1
\xi_{\mathbf{p_1}} -
i\left(\varepsilon_1+\omega\right)\xi_{\mathbf{p_1+q}}
-\xi_{\mathbf{p_1}}\xi_{\mathbf{p_1+q}}\right)}
{\left((\varepsilon-\varepsilon_1)^2 + (|\mathbf{p}|^2 -
2|\mathbf{p}||\mathbf{p_1}| z + |\mathbf{p_1}|^2) +
m^{2}_{\pi}\right)\left(\varepsilon^{2}_{1} +
\xi^{2}_{\mathbf{p_1}}\right)\left(\left(\varepsilon_{1} +
\omega\right)^{2} + \xi^{2}_{\mathbf{p_1+q}}\right)}.
\nonumber \\
\end{eqnarray}
Different from electron-phonon interacting systems, there is not any
peculiar energy scale that could be adopted to carry out approximate
analytical computations \cite{Migdal}. Thus the method used by
Migdal \cite{Migdal} cannot be applied to compute the above
integral. Below we calculate this integral as follows. An important
property of neutron matter is that only the neutrons excited around
the Fermi surface contribute to various physical quantities. Based
on this property, we assume that the two external neutron momenta of
the triangle diagram shown in Fig.~\ref{oneloopdiagram} are both
located at the Fermi surface, i.e., $|\mathbf{p}| = |\mathbf{p+q}| =
p_F$. The two external frequencies are taken as $\varepsilon =
\varepsilon+\omega = 0$. Under such approximations, we set
$\varepsilon=\omega=0$ and $|\mathbf{q}|\in [0, 2p_F]$. For
simplicity, we suppose that $|\mathbf{q}| = p_F$. We have verified
that the results are not visibly changed if $|\mathbf{q}|$ takes
other values around $p_F$. Now the ratio $\Gamma_{1}/\Gamma_{0}$ is
simplified to
\begin{eqnarray}
\frac{\Gamma_{1}}{\Gamma_{0}} &\approx&
\left(\frac{f_{\pi}}{m_{\pi}}\right)^2\int \frac{d\varepsilon_{1}
|\mathbf{p_1}|^2 d|\mathbf{p_1}| dz}{(2\pi)^3} \frac{\left(p^{2}_{F}
- 2p_{F}|\mathbf{p_1}|z + |\mathbf{p_1}|^2\right)
\left(\varepsilon^{2}_{1} - i\varepsilon_1
\left(\xi_{\mathbf{p_1}}+\xi_{\mathbf{p_1}+p_F}\right)-
\xi_{\mathbf{p_1}}\xi_{\mathbf{p_1}+p_F}\right)}{\left(
\varepsilon_{1}^{2} + (p^2_F - 2p_F |\mathbf{p_1}| z +
|\mathbf{p_1}|^2) + m^2_{\pi}\right)\left(\varepsilon^{2}_{1} +
\xi^{2}_{\mathbf{p_1}}\right)\left(\varepsilon^{2}_{1} +
\xi^{2}_{\mathbf{p_1}+p_F}\right)},\nonumber \\
&=& \left(\frac{f_{\pi}}{m_{\pi}}\right)^2\int^\Omega_{-\Omega}
\frac{d\varepsilon_1}{2\pi}\int^\Lambda_0\frac{|\mathbf{p_1}|^{2}
d|\mathbf{p_1}|}{2\pi}\int^{1}_{-1}\frac{dz}{2\pi}
\frac{\left(p^{2}_{F} - 2p_F|\mathbf{p_1}|z + |\mathbf{p_1}|^{2}
\right)\left(\varepsilon^{2}_{1} - \xi_{\mathbf{p_1}}
\xi_{\mathbf{p_1}+p_F}\right)}{\left(\varepsilon_{1}^{2} +
(p^{2}_{F} - 2p_F|\mathbf{p_1}|z + |\mathbf{p_1}|^2) +
m^2_{\pi}\right)\left(\varepsilon^{2}_{1} +
\xi^2_{\mathbf{p_1}}\right)\left(\varepsilon^{2}_{1} +
\xi^2_{\mathbf{p_1}+p_F}\right)}.\nonumber \\
\end{eqnarray}
This integral can be numerically computed. The numerical results are
presented in Fig.~\ref{ratio}. Apparently, the magnitude of the
ratio $\Gamma_{1}/\Gamma_{0}$ depends strongly on the neutron
density $\rho$. It is remarkable that the vertex correction is not
negligible even at very low densities. For instance, we find that
$\Gamma_{1}/\Gamma_{0} \approx 0.179$ at $\rho = 0.05\rho_{0}$ and
$\Gamma_{1}/\Gamma_{0} \approx 0.240$ at $\rho = 0.1\rho_{0}$. A
clear indication is that the contributions from vertex corrections
cannot be simply neglected.

It is necessary to estimate to what extend the values of $T_{c}$ and
$\rho_{c}$ are modified by the one-loop vertex correction. For
simplicity, we suppose that the contribution of one-loop vertex
correction can be effectively taken into account by replacing the
bare parameter $\Gamma_{0}$ with $\Gamma_{0}+\Gamma_{1}$. Then we
numerically solve the coupled equations of
$A_{0}(\varepsilon_n,|\mathbf{p}|)$ and
$\Delta_{s}(\varepsilon_n,|\mathbf{p}|)$ given by
Eqs.(\ref{eq:A0}-\ref{eq:Delta}) by making use of the new parameter
$\Gamma_{0}+\Gamma_{1}$. In Fig.~\ref{phasediagram}, we compare the
critical line obtained under ME (bare vertex) approximation to that
obtained by including the one-loop vertex correction. Obviously, the
one-loop vertex correction makes significant contributions to both
$\rho_{c}$ and $T_{c}$. We thus conclude that the ME theorem is
invalid in the present model and the widely used ME theory is not a
quantitatively reliable framework for the theoretical description of
the pion-mediated superfluid transition.

\section{Summary and Discussion \label{Sec:Summary}}

In summary, we apply the non-perturbative DS equation approach to
study the superfluid transition driven by the pion-mediated Cooper
pairing of neutrons. Based on a non-relativistic model of the
interaction between neutrons and $\pi^{0}$-mesons, we derive the
self-consistent integral equations of the renormalization function
$A_{0}(p)$ and the pairing function $\Delta_{s}(p)$. After
solving these two equations under the bare vertex approximation, we
obtained the energy-momentum dependence of these two functions,
extracted the critical temperature $T_c$ and the critical density
$\rho_{c}$, and also plotted a global phased diagram on $T$-$\rho$
plane. Then we went beyond bare vertex approximation and
incorporated the contributions of one-loop vertex correction into
the equations of $A_{0}(p)$ and $\Delta_{s}(p)$. We re-solved these
new equations and show that both $\rho_{c}$ and $T_{c}$ are
significantly altered by the vertex correction, implying the
invalidity of bare vertex approximation.

Our theoretical analysis need to be improved in three aspects. First
of all, it is certainly not sufficient to consider only the one-loop
vertex correction. Higher order vertex corrections should be taken
into account. As discussed in the last paragraph of
Sec.~\ref{Sec:DSE}, the approach developed in Refs.~\cite{Liu21,
Pan21} cannot be directly applied to precisely determine the full
vertex function for the $\pi N$ model. Nevertheless, we find it
still possible to properly generalize this approach to incorporate
the contributions of higher order vertex corrections. This work is
in progress and will be presented elsewhere \cite{Pan22}.
Furthermore, our model does not have any self-interaction term of
$\phi_{0}$ field. It would be interesting to examine the influence
of such an term as $\phi_{0}^{4}$ on the superfluid transition.
Another problem of our present work is that the $\pi N$ model
describes only the low density region with $\rho/\rho_{0} \leq 0.1$
and does not provide an adequate description of realistic neutron
stars. This is because exchanging pions only produces a long-range
attraction needed to form Cooper pairs but is not capable of
generating a short-range repulsion. For this reason, the pairing gap
shown in Fig.~\ref{compare} increases monotonously with growing
density $\rho$ within the range $\rho/\rho_{0} \leq 0.1$. In the
high density region (with $\rho/\rho_{0}>0.1$) where both long-range
attraction and short-range repulsion are important, the gap should
reach a maximum with growing $\rho$ and then tend to decrease as
$\rho$ further grows. Although BCS treatment neglects some important
effects, the phenomenological potential $V(r)$ used in BCS
calculations does include both the attraction and repulsion. Thus,
the pairing gap obtained by BCS calculations exhibits a maximum at
certain neutron density \cite{Lombardoreview, Sedrakian19}. To study
the neutron superfluid in the high density region with $\rho > 0.1
\rho_{0}$, we should couple neutrons to other mesons, such as
$\sigma$-meson, $\omega$-meson, and $\rho$-meson. Such an extended
model would be more realistic, but formally much more complicated
\cite{Walecka, Glendenningbook}. In a future project, we will
generalize the DS equation analysis to treat the multiple
neutron-meson couplings and to study both the $^{1}S_{0}$-wave and
$^{3}PF_{2}$-wave Cooper pairing instabilities within a broader
range of neutron density $\rho$.

\section*{ACKNOWLEDGEMENTS}

We would like to thank Xiao-Yin Pan and Xufen Wu for helpful
discussions. H.F.Z. is supported by the Natural Science Foundation
of China (Grants No. 12073026 and No. 11421303) and the Fundamental
Research Funds for the Central Universities.

\appendix

\section{Dyson-Schwinger equation of neutron propagator \label{sec:appa}}

One can derive rigorously the complete set of DS integral equations
of all the $n$-point correlation functions on the basis of the
Lagrangian of $\pi N$ interaction. Here we only provide the
derivational details that lead to the DS equation of neutron
propagator $G(p)$. Other DS equations can be derived in a similar
way \cite{Itzykson, Liu21, Pan21}.

Let us start from the following total Lagrangian density
\begin{eqnarray}
\mathcal{L}_T &=& \frac{1}{2}\widetilde{\psi}_n^{\dag}(x)
\left[i\partial_t - \xi_{\partial_\mathbf{x}}\widehat{A} ~\right]
\widetilde{\psi}_n(x) +
\frac{1}{2}\phi_0^{\dag}(x)
\mathbb{D}_{\phi_0}\phi^{}_0(x)
\nonumber \\
&& -\frac{i}{2}\frac{f_{\pi}}{m_{\pi}}\widetilde{\psi}_n^{\dag}(x)
\left[\widehat{C} \phi_0(x)\right]
\widetilde{\psi}_n(x)\nonumber \\
&& +J_{\phi_0}(x)\phi_0(x) +
\eta^{\dag}(x) \widetilde{\psi}_{n}(x)+
\widetilde{\psi}_{n}^{\dag}(x)\eta(x),
\end{eqnarray}
where $\mathbb{D}_{\phi_0}=-(\partial^2+m^2_{\pi})$, $J_{\phi_0}$,
$\eta^{{\dag}}$, and $\eta$ are external sources associated with
$\phi_0$, $\widetilde{\psi}_n$, and $\widetilde{\psi}_n^{{\dag}}$,
respectively. For simplicity, we define the following notations
\begin{eqnarray}
&&\widehat{A}=\lambda_3\otimes\sigma_0,\\
&& \widehat{C} = i\partial_i\widehat{C}_i = i\partial_x
\widehat{C}_x + i\partial_y\widehat{C}_y + i\partial_z\widehat{C}_z
= i\partial_x\lambda_0\otimes\sigma_1 + i\partial_y\lambda_3 \otimes
\sigma_2+i\partial_z\lambda_0\otimes\sigma_3.
\end{eqnarray}

To help the readers understand the calculational details, we first
list some basic rules of functional integration to be used later.
All the correlation functions are generated from three important
quantities: the partition function $Z[J_{\phi_0},\eta^{{\dag}},
\eta]$, the generating functional $W [J_{\phi_0},\eta^{{\dag}},
\eta]$ and the generating functional $\Xi [\phi_0,
\widetilde{\psi}_n, \widetilde{\psi}_n^{\dag}]$. They are defined as
follows:
\begin{eqnarray}
&&Z\left[J_{\phi_0},\eta^{{\dag}}, \eta\right]=\int \mathcal{D}\phi_0
\mathcal{D}\widetilde{\psi}_{n}^{\dag}
\mathcal{D}\widetilde{\psi}_n\exp{(i\int dtd^3\mathbf{x}
\mathcal{L}_T)},\\
&&W \left[J_{\phi_0},\eta^{{\dag}}, \eta\right]=-i\ln
Z\left[J_{\phi_0},\eta^{{\dag}}, \eta\right],\\
&&\Xi \left[\phi_0,\widetilde{\psi}_n,
\widetilde{\psi}_n^{\dag}\right]=W \left[J_{\phi_0},\eta^{{\dag}},
\eta\right]-\int\left(J_{\phi_0}\phi_0+\eta^{\dag} \widetilde{\psi}_n
+\widetilde{\psi}_n^{\dag} \eta\right).
\end{eqnarray}
The following identities will be frequently used:
\begin{eqnarray}
&&\frac{\delta W}{\delta J_{\phi_0}}=\langle\phi_0\rangle, \qquad
\frac{\delta W}{\delta \eta} =
-\langle\widetilde{\psi}_n^{\dag}\rangle, \qquad \frac{\delta
W}{\delta \eta^{\dag}}=\langle\widetilde{\psi}_n\rangle,\label{eq:A7}\\
&&\frac{\delta \Xi}{\delta \phi_0}=- J_{\phi_0}, \qquad\frac{\delta
\Xi}{\delta\widetilde{\psi}_n}= \eta^{\dag}, \qquad \frac{\delta
\Xi}{\delta \widetilde{\psi}_n^{\dag}}=-\eta.
\end{eqnarray}
$W[J_{\phi_0},\eta^{{\dag}}, \eta]$ generates all the connected
Green's functions and $\Xi [\phi_0,
\widetilde{\psi}_n,\widetilde{\psi}_n^{\dag}]$ generates all the
irreducible proper vertices of $\pi N$ coupling. For instance, the
full neutron propagator $G(x-x')$ and full pion propagator $F(x-x')$
are given by
\begin{eqnarray}
&&G(x-x')\equiv-i\langle\widetilde{\psi}_n(x)
\widetilde{\psi}^\dag_n(x')\rangle =
\frac{\delta^2W}{\delta\eta^\dag(x)\delta\eta(x')},\\
&&F(x-x')\equiv-i\langle\phi^{}_0(x)\phi^\dag_0(x')\rangle =
-\frac{\delta^2W}{\delta J_{\phi_0}(x)\delta J_{\phi_0}(x')},
\end{eqnarray}
and the $three$-point correlation function $\langle \phi_0
\widetilde{\psi}_n\widetilde{\psi}^\dag_n\rangle$ is defined as
follows:
\begin{eqnarray}
\langle\phi_0\widetilde{\psi}_n\widetilde{\psi}^\dag_n\rangle \equiv
\frac{\delta^3W}{\delta J_{\phi_0}\delta\eta^\dag \delta\eta}.
\end{eqnarray}
The following identity will also be frequently used:
\begin{eqnarray}
\frac{\delta^3W}{\delta J_{\phi_0}\delta\eta^\dag \delta\eta} =
-FG\frac{\delta^3\Xi}{\delta \phi_0\delta\widetilde{\psi}_n^{\dag}
\delta\widetilde{\psi}_n}G.
\label{eq:A12}
\end{eqnarray}

Now we derive the DS equation of the full neutron propagator. The
partition function $Z$ is invariant under an arbitrary infinitesimal
variation of spinor field $\widetilde{\psi}_n^{\dag}$, i.e.,
\begin{eqnarray}
\int \mathcal{D}\phi_0 \mathcal{D}\widetilde{\psi}_{n}^{\dag}
\mathcal{D}\widetilde{\psi}_n \frac{\delta}{i\delta
\widetilde{\psi}_n^{\dag}}\exp{(i\int dtd^3\mathbf{x}
\mathcal{L}_T)}=0.
\end{eqnarray}
It is easy to obtain
\begin{eqnarray}
\langle\left(i\partial_{t}-\xi_{\partial_\mathbf{x}}
\widehat{A}~\right)\widetilde{\psi}_n(x) - i \frac{f_{\pi}}{m_{\pi}}
\left[\widehat{C}\phi_0(x)\right]\widetilde{\psi}_n(x)+
2\eta(x)\rangle = 0.
\end{eqnarray}
Using the relations given by Eq.~(\ref{eq:A7}), we re-write the above
expression as
\begin{eqnarray}
-2\eta(x)&=&\left(i\partial_t -\xi_{\partial_\mathbf{x}} \widehat{A}
~\right)\frac{\delta W}{\delta \eta^{\dag}(x)} -
\frac{f_{\pi}}{m_{\pi}}\widehat{C}\frac{\delta^2 W}{\delta
J_{\phi_0}(x)\delta \eta^{\dag}(x)} - i\frac{f_{\pi}}{m_{\pi}}
\left(i\partial_i\frac{\delta W}{\delta J_{\phi_0}(x)}\right)
\widehat{C}_i\frac{\delta W}{\delta \eta^{\dag}(x)}.
\end{eqnarray}
The last term of the right-hand side (r.h.s.) vanishes upon removing
sources and can be directly omitted. Performing functional
derivative of both sides with respect to $\eta(x_2)$ and using the
identity (\ref{eq:A12}) leads to
\begin{eqnarray}
2\delta(x-x_{2}) &=& \left(i\partial_{t} - \xi_{\partial_\mathbf{x}}
\widehat{A}~\right)\frac{\delta^{2}W}{\delta\eta^{\dag}(x)\delta
\eta(x_2)} -\frac{f_{\pi}}{m_{\pi}}\widehat{C}\frac{\delta^3
W}{\delta J_{\phi_0}(z) \delta \eta^{\dag}(x)\delta \eta(x_2)}
\nonumber \\
&=& \left(i\partial_t - \xi_{\partial_\mathbf{x}}
\widehat{A}~\right)G(x-x_2)
\nonumber \\
&& +\frac{f_{\pi}}{m_{\pi}}\int dx_{1}dx_{3}dx_{4}
\widehat{C}F(x-x_3) G(x-x_1)\frac{\delta^3 \Xi}{\delta\phi_0(x_3)
\delta \widetilde{\psi}_n^{\dag}(x_1)\delta
\widetilde{\psi}_n(x_4)}G(x_4-x_2).
\end{eqnarray}
This expression can be re-cast into
\begin{eqnarray}
2G^{-1}(x-x_2) = \left(i\partial_{t} - \xi_{\partial_\mathbf{x}}
\widehat{A}~\right)\delta(x-x_2) +\frac{f_{\pi}}{m_{\pi}}\int dx_{1}
dx_{3}\widehat{C}F(x-x_3)G(x-x_1)
\Gamma_{\mathrm{v}}(x_3,x_1,x_2),\label{eq:A17}
\end{eqnarray}
where we have defined a truncated (without external legs) $\pi N$
interaction vertex function
\begin{eqnarray}
\Gamma_{\mathrm{v}}(x_3,x_1,x_2) = \frac{\delta^3 \Xi}{\delta
\phi_0(x_3) \delta \widetilde{\psi}_n^{\dag}(x_1)\delta
\widetilde{\psi}_n(x_2)}.
\end{eqnarray}
The full neutron propagator, the full pion propagator and
interaction vertex functions are Fourier transformed as
\begin{eqnarray}
G(x-\xi_1) &=& \int\frac{d^4p}{(2\pi)^4}e^{-ip(x-\xi_1)}G(p),\label{eq:A19}\\
F(x-\xi_2) &=& \int\frac{d^4q}{(2\pi)^4}e^{-iq(x-\xi_2)}F(q),\label{eq:A20}\\
\Gamma_{\mathrm{v}}(\xi_{1}-x, x-\xi_{2}) &=& \int\frac{d^4q
d^4p}{(2\pi)^8}e^{-i(p+q)(\xi_{1}-x)-ip(x-\xi_{2})}
\Gamma_{\mathrm{v}}(q,p).\label{eq:A21}
\end{eqnarray}
After making Fourier transformation Eqs.~(\ref{eq:A19}-\ref{eq:A21})
to Eq.~(\ref{eq:A17}), we eventually obtain the following DS
equation for the full neutron propagator:
\begin{eqnarray}
G^{-1}(p)=G_0^{-1}(p)-\frac{1}{2}\frac{f_{\pi}}{m_{\pi}}\int
\frac{d^4q}{(2\pi)^4} \widehat{C}_1 G(p+q)
F(q)\Gamma_{\mathrm{v}}(q,p),
\end{eqnarray}
where
\begin{eqnarray}
\widehat{C}_1 &=& q_x\lambda_0\otimes\sigma_1 + q_y\lambda_3
\otimes\sigma_2+q_z\lambda_0\otimes\sigma_3.
\end{eqnarray}

The DS equation of the full pion propagator can be derived in an
analogous way. We will not give the derivational details and only
present their final expression
\begin{eqnarray}
F^{-1}(q) = F_{0}^{-1}(q) + \frac{1}{2}\frac{f_{\pi}}{m_{\pi}}\int
\frac{d^4p}{(2\pi)^4} \mathrm{Tr} \left[\widehat{C}_1
G(p+q)\Gamma_{\mathrm{v}}(q,p)G(q)\right].\label{eq:A24}
\end{eqnarray}

We next derive an exact relation that connects $F(q)$ and $F_{0}(q)$
with $\Gamma_{\mathrm{v}}(q,p)$. It is clearly true that $Z$ is not
changed by infinitesimal variations of pion field $\phi_{0}(x)$,
which indicates that
\begin{eqnarray}
\langle\mathbb{D}_{\phi_{0}} \phi_{0}(x) + i\frac{1}{2}
\frac{f_{\pi}}{m_{\pi}} \left[i\partial_{i}
\widetilde{\psi}_{n}^{\dag}(x) \widehat{C}_{i}
\widetilde{\psi}_{n}(x)\right]+ J_{\phi_0}(x)\rangle=0.
\end{eqnarray}
This equation can be converted to
\begin{eqnarray}
-i\frac{1}{2}\frac{f_{\pi}}{m_{\pi}}\langle \left[i\partial_{i}
\widetilde{\psi}_{n}^{\dag}(x)\widehat{C}_i
\widetilde{\psi}_n(x)\right]\rangle = \mathbb{D}_{\phi_{0}}
\frac{\delta W}{\delta J_{\phi_{0}}(x)}+J(x),
\end{eqnarray}
which, after taking the functional derivative with respect to
$\eta^{\dag}$ and $\eta$ in order, leads to
\begin{eqnarray}
&&\frac{\delta^2}{\delta\eta^{\dag}(x_1)\delta\eta(x_2)}
\langle\left[i\partial_i\widetilde{\psi}_n^{\dag}(x)
\widehat{C}_i\widetilde{\psi}_n(x)\right]\rangle
\nonumber \\
&=& \langle\left[i\partial_i\widetilde{\psi}_n^{\dag}(x)
\widehat{C}_i \widetilde{\psi}_n(x)\right]\widetilde{\psi}_n(x_1)
\widetilde{\psi}_n^{\dag}(x_2)\rangle
\nonumber \\
&=& \langle \left[i\partial_x\widetilde{\psi}_n^{\dag}(x)
\widehat{C}_x \widetilde{\psi}_n(x) +
i\partial_y\widetilde{\psi}_n^{\dag}(x)\widehat{C}_y
\widetilde{\psi}_n(x)+
i\partial_z\widetilde{\psi}_n^{\dag}(x)\widehat{C}_z
\widetilde{\psi}_n(x)\right]\widetilde{\psi_n}(x_1)
\widetilde{\psi}_n^{\dag}(x_2)\rangle
\nonumber \\
&=& i2\left(\frac{f_{\pi}}{m_{\pi}}\right)^{-1}
\mathbb{D}_{\phi_0}\frac{\delta^3 W}{\delta J_{\phi_{0}}(x)
\delta\eta^{\dag}(x_1)\delta\eta(x_2)}.
\end{eqnarray}
Making use of the three current vertex functions defined by
Eq.~(\ref{eq:currentvertex}) and the identity (\ref{eq:A12}),
we obtain the following exact relation
\begin{eqnarray}
&& \int dx_{3}dx_{4}G(x_1-x_3)\left[i\partial_x
\Gamma_{\widehat{C}_x}(x,x_3,x_4) + i\partial_y
\Gamma_{\widehat{C}_y}(x,x_3,x_4) + i\partial_z
\Gamma_{\widehat{C}_z}(x,x_3,x_4)\right]G(x_4-x_2) \nonumber \\
&=& \int dx_{3}dx_{4} dx_{5} i 2\left(\frac{f_{\pi}}{m_{\pi}}
\right)^{-1} \mathbb{D}_{\phi_{0}}F(x-x_5)G(x_1-x_3)
\Gamma_{\mathrm{v}}(x_5,x_3,x_4)G(x_4-x_2).
\label{eq:exactrelationrealspace}
\end{eqnarray}
The current vertex functions are Fourier transformed
as
\begin{eqnarray}
\Gamma_{\widehat{C}_{x,y,z}}(\xi_{1}-x, x-\xi_{2}) &=&
\int\frac{d^4q d^4p}{(2\pi)^{8}}e^{-i(p+q)(\xi_{1}-x)-ip(x-\xi_{2})}
\Gamma_{\widehat{C}_{x,y,z}}(q,p).\label{eq:A29}
\end{eqnarray}
Making Fourier transformation to
Eq.~(\ref{eq:exactrelationrealspace}) gives rise to
\begin{eqnarray}
i2\left(\frac{f_{\pi}}{m_{\pi}}\right)^{-1}F_{0}^{-1}(q)
F(q)\Gamma_{\mathrm{v}}(q,p) = q_{x}\Gamma_{\widehat{C}_x}(q,p) +
q_y \Gamma_{\widehat{C}_y}(q,p) + q_z \Gamma_{\widehat{C}_z}(q,p).
\end{eqnarray}
It is more convenient to re-write this relation as
\begin{eqnarray}
F(q)\Gamma_{\mathrm{v}}(q,p) = -i\frac{1}{2} \frac{f_{\pi}}{m_{\pi}}
F_{0}(q)\left[q_{x}\Gamma_{\widehat{C}_x}(q,p) + q_y
\Gamma_{\widehat{C}_{y}}(q,p) + q_{z}
\Gamma_{\widehat{C}_z}(q,p)\right].
\end{eqnarray}
This relation is used in Sec.~\ref{Sec:DSE}.


\begin{thebibliography}{99}

\bibitem{Itzykson}
C. Itzykson and J.-B. Zuber, \emph{Quantum Field Theory}
(McGraw-Hill, New York, 1980).

\bibitem{Walecka}
J. D. Walecka, \emph{Theoretical Nuclear and Subnuclear Physics},
(Oxford University Press, 1995).

\bibitem{AGD}
A. A. Abrikosov, L. P. Gor'kov, and I. Y. Dzyaloshinskii,
\emph{Quantum Field Theoretical Methods in Statistical Physics}
(Pergamon Press Inc., 1965).

\bibitem{BCS}
J. Bardeen, L. N. Cooper, and J. R. Schrieffer, Phys. Rev. {\bf
108}, 1175 (1957).

\bibitem{Migdal60}
A. B. Migdal, Soviet Physics JETP {\bf 10}, 176 (1960).

\bibitem{Roberts00}
C. D. Roberts and S. M. Schmidt, Prog. Part. Nucl. Phys. {\bf 45},
S1-S103 (2000).

\bibitem{Eliashberg}
G. M. Eliashberg, Sov. Phys. JETP {\bf 11}, 696 (1960).

\bibitem{Migdal}
A. Migdal, Sov. Phys. JETP {\bf 7}, 996 (1958).

\bibitem{Scalapino}
D. J. Scalapino, \emph{The electron-phonon interaction and
strong-coupling superconductivity}, in \emph{Superconductivity},
edited by R. D. Parks (Marcel Dekker, Inc. New York, 1969).

\bibitem{Terasaki02}
J. Terasaki, F. Barranco, R. A.
Broglia, E. Vigezzi, and P. F. Bortignon, Nucl. Phys. A {\bf697},
127 (2002).

\bibitem{Sedrakian98}
A. Sedrakian, Astrophys. \& Space
Sci. {\bf236}, 267 (1996); A. Sedrakian, in \emph{Proceedings of the
International Workshop Hirshegg '98, Nuclear Astrophysics}, edited
by M. Buballa, W. N\"{o}renberg, J. Wambach, and A. Wirzba (GSI,
Darmstadt, 1998), p.~54, astro-ph/9801239.

\bibitem{Fernandes}
M. N. Gastiasoro, J. Ruhman, and R. M. Fernandes, Ann. Phys. {\bf
417}, 168107 (2020).

\bibitem{Liu21}
G.-Z. Liu, Z.-K. Yang, X.-Y. Pan, and J.-R. Wang, Phys. Rev. B {\bf
103}, 094501 (2021).

\bibitem{Margueron}
J. Margueron, H. Sagawa, and K. Hagino, Phys. Rev. C {\bf 76},
064316 (2007).

\bibitem{Mao}
S. Mao, X. Huang, and P. Zhuang, Phys. Rev. C {\bf 79}, 034304
(2009).

\bibitem{Sun10}
B. Y. Sun, H. Toki, and J. Meng, Phys. Lett. B {\bf 134}, 683
(2010).

\bibitem{Sun}
T. T. Sun, B. Y. Sun, and J. Meng, Phys. Rev. C {\bf 86}, 014305
(2012).

\bibitem{Stein}
M. Stein, A. Sedrakian, X.-G. Huang, and J. W. Clark, Phys. Rev. C
{\bf 90}, 065804 (2014).

\bibitem{Glendenningbook}
N. K. Glendenning, \emph{Compact Stars} (Springer, New York, 2000).

\bibitem{Shapiro}
S. L. Shapiro and S. A. Teukolsky, \emph{Black Holes, White Dwarfs,
and Neutron Stars: The Physics of Compact Objects} (WILEY-VCH Verlag
GmbH $\&$ Co. KGaA, 2004).

\bibitem{Pagereview}
D. Page, J. M. Lattimer, M. Prakash, and A. W. Steiner,
\emph{Pairing and superfluidity of nucleons in neutron stars}, in
\emph{Novel Superfluids}, edited by K.H. Bennemann, J.B. Ketterson
(Oxford University Press, Oxford, UK, 2013).

\bibitem{Lombardoreview}
U. Lombardo and H.-J. Schulze, in \emph{Lecture Notes in Physics}
(Springer, New York, 2001), Vol.~578.

\bibitem{Sedrakian19}
A. Sedrakian and J. W. Clark, Eur. Phys. J. A {\bf 55}, 167 (2019).

\bibitem{Baldo10}
M. Baldo, U. Lombardo, S. S. Pankratov, and E. E. Saperstein, J.
Phys. G: Nucl. Part. Phys. {\bf 37}, 064016 (2010).

\bibitem{Baldo98}
M. Baldo, {\O}. Elgar{\o}y, L. Engvik, M. Hjorth-Jensen, and H.-J.
Schulze, Phys. Rev. C {\bf 58}, 1921 (1998).

\bibitem{Elgaroy98}
{\O}. Elgar{\o}y and M. Hjorth-Jensen, Phys. Rev. C {\bf 57}, 1174
(1998).

\bibitem{Walecka74}
J. D. Walecka, Ann. Phys. {\bf 83}, 491 (1974).

\bibitem{Hirose07}
S. Hirose, M. Serra, P. Ring, T. Otsuka, and Y. Akaishi, Phys. Rev.
C {\bf 75}, 024301 (2007).

\bibitem{Kucharek91}
H. Kucharek and P. Ring, Z. Phys. A {\bf 339}, 23 (1991).

\bibitem{Boguta77}
J. Boguta, and A. R. Bodmer, Nucl. Phys. A {\bf 292}, 413 (1977)

\bibitem{Muller96}
H. M\"{u}ller, and B. D. Serot, Nucl. Phys. A {\bf 606}, 508 (1996)

\bibitem{Baade34}
W. Baade and F. Zwicky, Phys. Rev. {\bf 45}, 138 (1934).

\bibitem{Gold68}
T. Gold, Nature {\bf 218}, 731 (1968).

\bibitem{Hewish68}
A. Hewish, S. J. Bell, J. D. H. Pilkington, P. F. Scott, and R. A.
Collins, Nature {\bf 217}, 709 (1968).

\bibitem{Baym69}
G. Baym, C. Pethick, D. Pines, and M. Ruderman, Nature {\bf 224},
872 (1969).

\bibitem{Ho09}
W. C. G. Ho and C. O. Heinke, Nature (London) {\bf 462}, 71 (2009).


\bibitem{Nambu60}
Y. Nambu, Phys. Rev. {\bf 117}, 648 (1960).

\bibitem{Weise}
T. Ericson and W. Weise, \emph{Pions and Nuclei} (Claredon, Oxford,
1988).

\bibitem{Sedrakian03}
A. Sedrakian, Phys. Rev. C {\bf 68}, 065805 (2003).

\bibitem{Pan21}
X.-Y. Pan, Z.-K. Yang, X. Li, and G.-Z. Liu, Phys. Rev. B {\bf 104},
085141 (2021).

\bibitem{Serra01}
M. Serra, A. Rummel, and P. Ring, Phys. Rev. C {\bf 65}, 014304
(2001).

\bibitem{Kucharek89}
H. Kucharek, P. Ring, P. Schuck, R. Bengtsson, and M. Girod, Phys.
Lett. B {\bf 216}, 249 (1989).

\bibitem{Furtado22}
U. J. Furtado, S. S. Avancini, and J. R. Marinelli, J. Phys. G:
Nucl. Part. Phys. {\bf 49}, 025202 (2022).

\bibitem{Erkelenz69}
K. Erkelenz, K. Holinde, and K.
Bleuler, Nucl. Phys. A {\bf139}, 308 (1969).

\bibitem{Erkelenz74}
K. Erkelenz, Phys. Rep. {\bf 13},
191 (1974).

\bibitem{Holinde87}
R. Machleidt, K. Holinde, and Ch.
Elster, Phys. Rep. {\bf 149}, 1 (1987).

\bibitem{Erkelenz71}
K. Erkelenz, R. Alzetta, and K.
Holinde, Nucl. Phys. A {\bf176}, 413 (1971).

\bibitem{Holinde72}
K. Holinde, K. Erkelenz, and R.
Alzetta, Nucl. Phys. A {\bf194}, 161 (1972).


\bibitem{Pan22}
X.-Y. Pan, H.-F. Zhu, and G.-Z. Liu, in preparation (2022).

\end{thebibliography}
\end{document}